\documentclass{article}

\usepackage{arxiv}

\usepackage[utf8]{inputenc} 
\usepackage[T1]{fontenc}    
\usepackage{hyperref}       
\usepackage{url}            
\usepackage{booktabs}       
\usepackage{amsfonts}       
\usepackage{nicefrac}       
\usepackage{microtype}      
\usepackage{lipsum}
\usepackage{graphicx}

\usepackage{textcomp}
\usepackage{xcolor}
\usepackage{float}
\usepackage{subfig}
\usepackage{xspace}

\usepackage{multirow}
\usepackage{pifont}
\usepackage{algorithm}
\usepackage{algpseudocode}
\usepackage{threeparttable}
\usepackage{mathtools}
\usepackage{siunitx}
\usepackage{tikz}

\newcommand{\figref}[1]{\figurename~{\ref{#1}}}
\newcommand{\tabref}[1]{Table~{\ref{#1}}}

\newcommand{\eqnref}[1]{Equation~{(\ref{#1})}}

\newcommand{\type}[1]{\texttt{#1}}
\newcommand*{\circled}[1]{\lower.7ex\hbox{\tikz\draw (0pt, 0pt)%
    circle (.45em) node {\makebox[.45em][c]{\small #1}};}}
\DeclarePairedDelimiter\floor{\lfloor}{\rfloor}

\newcommand{\aname}{TATAA\xspace}

\title{\aname: Programmable Mixed-Precision Transformer Acceleration with a Transformable Arithmetic Architecture}

\author{
 Jiajun Wu$^*$ \\
  Department of Electrical and Electronic Engineering\\
  University of Hong Kong\\
  Hong Kong \\
  \texttt{jjwu@eee.hku.hk} \\
   \And
 Mo Song$^*$ \\
  Department of Electrical and Electronic Engineering\\
  University of Hong Kong\\
  Hong Kong \\
  \texttt{songmo@eee.hku.hk} \\
  \And
 Jingmin Zhao \\
  Department of Electrical and Electronic Engineering\\
  University of Hong Kong\\
  Hong Kong \\
  \texttt{jmzhao@eee.hku.hk} \\
  \And
 Yizhao Gao \\
  Department of Electrical and Electronic Engineering\\
  University of Hong Kong\\
  Hong Kong \\
  \texttt{yzgao@eee.hku.hk} \\
  \And
 Jia Li \\
  Department of Electrical and Electronic Engineering\\
  University of Hong Kong\\
  Hong Kong \\
  \texttt{lijia@eee.hku.hk} \\
  \And
 Hayden Kwok-Hay So \\
  Department of Electrical and Electronic Engineering\\
  University of Hong Kong\\
  Hong Kong \\
  \texttt{hso@eee.hku.hk} \\
  \And
}

\begin{document}
\maketitle
\def\thefootnote{*}\footnotetext{These authors contributed equally to this work}\def\thefootnote{\arabic{footnote}}

\begin{abstract}
Modern transformer-based deep neural networks present unique technical challenges for effective acceleration in real-world applications. 
Apart from the vast amount of linear operations needed due to their sizes, modern transformer models are increasingly reliance on precise non-linear computations that make traditional low-bitwidth quantization methods and fixed-dataflow matrix accelerators ineffective for end-to-end acceleration.
To address this need to accelerate both linear and non-linear operations in a unified and programmable framework, this paper introduces \aname.
\aname employs 8-bit integer (\type{int8}) arithmetic for quantized linear layer operations through post-training quantization, while it relies on \type{bfloat16} floating-point arithmetic to approximate non-linear layers of a transformer model.
\aname hardware features a transformable arithmetic architecture that supports both formats during runtime with minimal overhead, enabling it to switch between a systolic array mode for \type{int8} matrix multiplications and a SIMD mode for vectorized \type{bfloat16} operations.
An end-to-end compiler is presented to enable flexible mapping from  emerging transformer models to the proposed hardware.
Experimental results indicate that our mixed-precision design incurs only \SI{0.14}{\percent} to \SI{1.16}{\percent} accuracy drop when compared with the pre-trained single-precision transformer models across a range of vision, language, and generative text applications. Our prototype implementation on the Alveo U280 FPGA currently achieves \num{2935.2} GOPS throughput on linear layers and a maximum of \num{189.5} GFLOPS for non-linear operations, outperforming related works by up to $1.45\times$ in end-to-end throughput and $2.29\times$ in DSP efficiency, while achieving $2.19\times$ higher power efficiency than modern NVIDIA RTX4090 GPU.

\end{abstract}
\section{Introduction}
Since its introduction in 2017, the Transformer model \cite{vaswani2017attention} and its variations have rapidly risen to the forefront of modern deep learning architectures. 
Unlike previous-generation convolutional neural networks (CNNs) that were based predominantly on linear operations, modern transformer models are increasingly reliance on \emph{high-precision non-linear operations} in their designs.
For instance, the self-attention mechanism of a transformer model is typically based on the SoftMax function, which has been demonstrated to require high precision computation in order to achieve a model's accuracy~\cite{softermax}.
Normalization layers such as LayerNorm \cite{ba2016layer} or root mean square normalization (RMSNorm)~\cite{zhang-sennrich-neurips19}, require complex nonlinear operations on data that cannot easily be fused into preceding linear layers.
Finally, sophisticated activation functions such as GELU \cite{hendrycks2016gaussian}, SiLU \cite{ramachandran2017searching} and SwiGLU \cite{shazeer2020glu} are often used in transformer models which require precise computation, unlike in CNNs.

\begin{figure}[tbp]
    \centering
    \includegraphics[width=\linewidth]{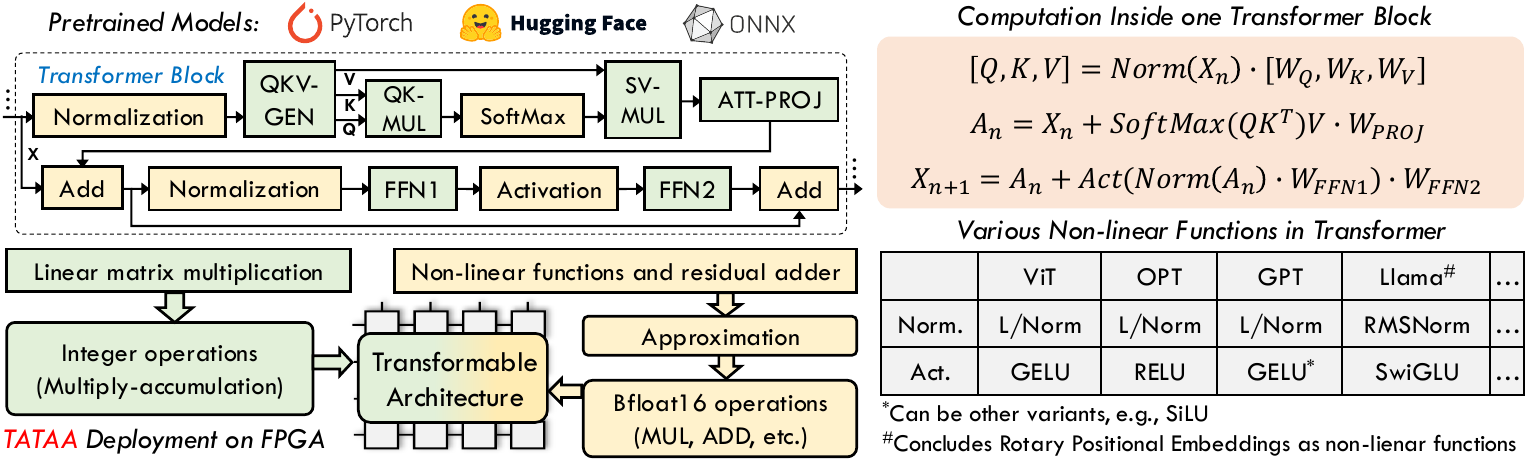}
    \caption{Illustration of a typical Transformer block, and how \aname maps different operations in Transformers including linear \textit{MatMul} and a variety of non-linear functions into transformable architecture, based on a general end-to-end framework design.}
    \label{fig:bg-trans}
\end{figure}

To address the need to approximate these non-linear functions with high precision and high performance, specialized hardware modules have previously been extensively explored~\cite{sze2017efficient, lin2018approximation, geng2019hardware, Zeng2024FlightLLMEL}. Yet, customized hardware must be designed for every non-linear function that is being employed in a model, which is impractical when new non-linear operations for transformers are still actively being developed in this rapidly-evolving field~\cite{touvron2023llama,su2024roformer}.
Other researchers have focused on quantizing such non-linear functions into low-bitwidth fixed point formats in order to reduce the computation complexity~\cite{li2023vit, kim2021bert, Marchisio2023SwiftTron, khannpe}. Due to the outliers in transformers~\cite{zadeh2022mokey, wei2022outlier}, retraining is generally required for such quantization to maintain good accuracy. However, the large size of modern transformer models, data availability and privacy concerns, have all made such retraining method impractical in most real-world scenarios.
Besides, existing accelerators either rely on individual and specific non-linear function units \cite{wang2022via, huang2023integer, bai2023fet}, or attempt to handle non-linear functions with general arithmetic logic units \cite{khannpe}. Both strategies often lead to increased hardware overhead, reduced hardware efficiency, and it complicates the workload balance between linear and non-linear layers.
Instead, to improve future compatibility, a general-purpose transformer accelerator that can be reprogrammed to support new non-linear functions in floating-point arithmetic with low hardware overhead is highly desirable.


In this paper, we present \aname, a novel end-to-end framework for flexible and quantized \underline{T}ransformer \underline{A}cceleration using a \underline{T}rans\-form\-able \underline{A}rithmetic \underline{A}rchitecture that supports both floating-point and integer operations (\figref{fig:bg-trans}). In \aname, static post-training quantization (PTQ) is performed on the linear layers of a transformer model to facilitate \num{8}-bit integer (\type{int8}) operations. On the other hand, non-linear layers are performed using high-precision \type{bfloat16} operations. In contrast to certain previous efforts that reduced data bitwidths in nonlinear layers, we alleviate the need for retraining by maintaining high bitwidth data formats in these layers, all the while ensuring high efficiency during execution.
To support both \type{int8} and \type{bfloat16} operations efficiently, \aname hardware architecture consists of dual-mode processing units (DMPUs) that feature configurable arrays of integer processing elements (PE). The proposed architecture can be transformed between a systolic array for \type{int8} matrix multiplications and a SIMD-like vectorized \type{bfloat16} computing unit. In particular, the proposed \aname architecture employs a single type of processing unit, which is reused for all run-time operations, leveraging the bit field patterns of \type{bfloat16}. This design choice minimizes hardware overhead and maximizes flexibility compared to previous studies.  By minimizing the overhead for run-time reconfiguration, the proposed transformable architecture ensures the high hardware processing density necessary to deliver the highest performance on FPGAs with limited resources.  Finally,  a compilation framework is developed that maps the user-provided transformer models to the custom instruction set architecture (ISA) of the TATAA processor cores to facilitate all operations in both linear and non-linear layers.

To the best of our knowledge, \aname is the first FPGA-based acceleration framework for transformer inference that integrates floating-point non-linear functions into integer-based linear processing units. It is programmable and is ready to support emerging transformer models with potentially new non-linear functions. Our experimental results indicate that when simulating model performance with a hybrid data format for transformer inference, \aname achieves only a minimal accuracy reduction, with \SI{0.14}{\percent} to \SI{1.16}{\percent} decrease across all evaluated models compared to the original pre-trained model in single-precision floating point (\type{fp32}).  Additionally, the FPGA accelerator reaches a peak throughput of \num{2935.2} giga-operations-per-second (GOPS) for \type{int8} linear operations and a peak throughput of \num{169.8} giga-floating-point-operations-per-second (GFLOPS) when the processor is configured for \type{bfloat16} non-linear operations at a clock frequency of \SI{225}{\mega\hertz}. Compared to related studies, \aname achieves up to $1.45\times$ higher throughput and $2.29\times$ higher throughput efficiency on DSP blocks. With the transformable architecture for non-linear functions, our implementation achieves $4.25\times$ lower latency for these complex \type{bfloat16} operations compared with other works, while supporting flexible and general compilation for emerging functions. Our end-to-end compilation framework also presents optimal mapping from non-linear functions to hardware ISA by appropriate approximation schemes and efficient dataflow control. Moreover, compared to state-of-the-art GPUs, our \aname architecture outperforms a maximum $2.19\times$ higher power efficiency over a variety of transformer models. This prototype underscores the potential of the \aname approach for expediting transformer models and sets the stage for future optimization at the microarchitectural level, while our extensive compilation flow opens up a significant optimization space for non-linear functions and to quickly adapt to future transformer-based model as they are being developed.

\section{Background \& Related Works}

\subsection{Transformer and Quantization}

The Transformer architecture \cite{vaswani2017attention}, along with its various derivatives, such as the Vision Transformer (ViT) \cite{dosovitskiy2020image, touvron2021training, liu2021swin}, and language models including BERT \cite{devlin2018bert}, OPT \cite{zhang2022opt}, GPT \cite{zhang2022opt}, and Llama \cite{touvron2023llama}, have been extensively utilized in numerous applications. Regardless of the overall topology, the fundamental unit in these models, the transformer block, typically includes components such as MLP, activation, etc. \figref{fig:bg-trans} illustrates a block in ViT, where the green components represent linear matrix multiplications (\textit{MatMul}), and the yellow components denote non-linear functions or residual adders. In detail, linear \textit{MatMul} layers encompass the generation of self-attention entries (QKV-GEN), multiplication of $QK^{T}$ to calculate attention weights (QK-MUL), the \textit{MatMul} operation between the SoftMax-applied attention scores and $V$ entries (SV-MUL), followed by three feed-forward networks (ATT-PROJ, FFN1, and FFN2). Accordingly, the non-linear functions incorporate normalization, SoftMax, and activation, which differ across various transformer-based models, as illustrated in \figref{fig:bg-trans}.


Due to the efficiency of integer \textit{MatMul} operations on hardware, linear quantization has been widely applied in modern deep neural networks (DNNs), including transformer models, to reduce memory footprint and computational complexity. \eqnref{eq:quant} presents the switching between floating-point ($fp$) format and quantized integer number ($q$), in terms of a basic multiplication $z=x\cdot y$. Such a quantized basic operation can be extended to any linear operations (e.g., \textit{MatMul}) by giving a sufficient intermediate integer bitwidth to avoid overflow.

\begin{equation}\label{eq:quant}
\begin{gathered}
    q_{x} = \floor{x_{fp}/S_{x}}, q_{y} = \floor{y_{fp}/S_{y}} \\
    z_{fp} = x_{fp} \cdot y_{fp} = q_{x}S_{x} \cdot q_{y}S_{y} \\
    q_{z} = \floor{q_{x}S_{x} \cdot q_{y}S_{y}/S_{z}} = \floor{(q_{x}\cdot q_{y})S_{x}S_{y}/S_{z}}
\end{gathered}
\end{equation}

According to \eqnref{eq:quant}, the key components to deploy quantized operations are scaling factors. To determine the scaling factors for each layer in transformer models, the primary quantization approaches are post-training quantization (PTQ) \cite{bai2022towards, liu2021post, lin2022fqvit, zadeh2022mokey} and quantization-aware training (QAT) \cite{li2023vit, kim2021bert, dettmers2023qlora}. Since QAT requires fine-tuning and retraining with expensive overhead \cite{nagel2021white}, exploring the static PTQ approach is more practical in transformer applications and is applied in our quantization framework \cite{xiao2023smoothquant, lin2022fqvit}. In \aname, we develop the quantization emulator based on a hardware matching style instead of \textit{fake quantization} to get more convincing results, following the HAWQ setups \cite{yao2021hawq}. Besides, \aname can integrate existing PTQ schemes like FQ-ViT \cite{lin2022fqvit} and SmoothQuant \cite{xiao2023smoothquant}. The static PTQ scheme only requires to access a relatively small part of dataset for calibration and getting all the scaling factors (i.e., $S_{x}, S_{y}, S_{z}$) before deploying inference. Once we have the scaling factors, our mixed-precision quantization can be done through \eqnref{eq:quant}, switching between floating-point and integer numbers for different kinds of layers.

\subsection{Non-linear Functions in Transformer}

Beyond integer-based \textit{MatMul} layers, transformers require non-linear functions to achieve high performance. For example, SoftMax \cite{bridle1989training}, Normalization (e.g., LayerNorm, RMSNorm) \cite{ba2016layer, zhang-sennrich-neurips19}, and activation functions (e.g., GELU\footnote{We use $tanh$ approximation of GELU function in this work}, SiLU, SwiGLU) \cite{hendrycks2016gaussian, ramachandran2017searching, shazeer2020glu} in \eqnref{eq:non-linear}, are commonly used in transformers to extract self-attention features, activate the feed-forward block, and normalize the output of each block, respectively. These non-linear operations and their variants are essential yet costly building basis of transformer models and can be difficult to implement directly or efficiently on hardware. The linear quantization methods described in \eqnref{eq:quant} no longer suit non-linear functions due to more complex operations and higher range \& precision requirement during runtime.


\begin{equation}\label{eq:non-linear}
\begin{gathered}
\operatorname{SoftMax}\left(\mathbf{x}\right)=\frac{\exp \left(\mathbf{x}\right)}{\sum_i \exp \left(x_i\right)} \\
\operatorname{LayerNorm}(\mathbf{x})=\frac{\mathbf{x}-\mathrm{E}[\mathbf{x}]}{\sqrt{\operatorname{Var}[\mathbf{x}]+\epsilon}} \cdot \gamma+\beta, \quad \operatorname{RMSNorm}(\mathbf{x})=\frac{\mathbf{x}}{\sqrt{\mathrm{E}[\mathbf{x^2}]}}\cdot \gamma \\
\operatorname{GELU}(\mathbf{x})=0.5\cdot\mathbf{x}\cdot\left(1+\operatorname{tanh}\left(\sqrt{2 / \pi}\left(\mathbf{x}+0.044715\mathbf{x}^3\right)\right) \right) \\
\operatorname{SiLU}(\mathbf{x}) = \mathbf{x} \cdot \sigma(\mathbf{x}) = \mathbf{x} \cdot \frac{1}{1+\exp(-\mathbf{x})}, \quad \operatorname{SwiGLU}(\mathbf{x}) = \left( \mathbf{x} \cdot \sigma(\mathbf{x}) \right) \cdot \left( \mathbf{x} \cdot \sigma(W\mathbf{x} + b) \right)
\end{gathered}
\end{equation}





To alleviate hardware inefficiency, several works proposed approximation techniques based on integer-only arithmetic \cite{kim2021bert} or reduced precision computation \cite{DBLP:journals/corr/abs-2103-09301}, and several researchers argued that LUT-based methods \cite{han2017ese, dong2019hardware} demonstrated both negligible model accuracy degradation and higher computational efficiency. Especially in \cite{Zeng2024FlightLLMEL}, the miscellaneous non-linear operations are element-wise handled by a special function unit, which requires special breakpoints of vectors to perform in fine granularity to hide computation latency and wire resources. LogicNets\cite{umuroglu2020logicnets} and NullaNet \cite{nazemi2018nullanet} act as general architectures that encapsulate all the operations embedded in linear or non-linear layers by enumerating the truth table values and can be further optimized following logic optimization algorithms. All these implementations of non-linear functions necessitate additional hardware units beyond the linear processing units with larger bitwidth support. In \aname, we opt to utilize the same hardware processing units for both types of layers and comprehensive support.


\subsection{Transformer Accelerators}

\begin{table}[tbp]
\centering
\caption{Qualitative Comparison with Related Software-Hardware Co-Design Transformer Acceleration Frameworks}
\resizebox{0.92\linewidth}{!}{
\begin{tabular}{@{}ccccccc@{}}
\toprule
Work & \begin{tabular}[c]{@{}c@{}}End-to-End\\ Support\end{tabular} & \begin{tabular}[c]{@{}c@{}}Retrain/\\ Fine-Tuning\end{tabular} & \begin{tabular}[c]{@{}c@{}}Hardware \\ Platform\end{tabular} & \begin{tabular}[c]{@{}c@{}}\textit{MatMul}\\ Data Format\end{tabular} & \begin{tabular}[c]{@{}c@{}}Non-linear\\ Data Format\end{tabular} & \begin{tabular}[c]{@{}c@{}}Non-linear\\ Implementation\end{tabular} \\ \midrule
$A^3$ \cite{hamA3}& No & N/A & ASIC & \type{int8} & Integer & N/A \\ \midrule
Mokey \cite{zadeh2022mokey}& No & No & ASIC & \type{fxp} & N/A & N/A \\ \midrule
Auto-ViT-Acc \cite{lit2022auto}& No & Yes & FPGA & Mixed Integer & \type{fp32} & Host CPU \\ \midrule
Zhang et al. \cite{zhang2021algorithm}& No & Yes & FPGA & \type{int8} & \type{fp32} & N/A \\ \midrule
FQ-BERT \cite{liu2021hardware}& Yes & Yes & FPGA & \type{int8} & \type{fxp} & Special Units \\ \midrule
I-ViT \cite{li2023vit}& Yes & Yes & GPU & \type{int8} & Integer & GPU Vector Units \\ \midrule
I-BERT \cite{kim2021bert} & Yes & Yes & GPU & \type{int8} & Integer & GPU Vector Units \\ \midrule
Transformer Engine \cite{trans-engine} & Yes & N/A & GPU & \type{fp8} & \type{fp16}/\type{fp32} & GPU Vector Units \\ \midrule
ViA \cite{wang2022via}& Yes & No & FPGA & \type{fp16} & \type{fp16} & Special Units \\ \midrule
SwiftTron \cite{Marchisio2023SwiftTron}& Yes & Yes & ASIC & \type{int8} & \type{fxp} & Special Units \\ \midrule
FTRANS \cite{ftrans}& Yes & No & FPGA & \type{fp16} & \type{fp32} & Special Units \\ \midrule
Huang et al. \cite{huang2023integer} & Yes & Yes & FPGA & \type{int8} & \type{int8} & Special Units \\ \midrule
FlexRun \cite{flexrun} & Yes & Yes & FPGA & \type{int8} & \type{fp32} & Vector Units \\ \midrule
SSR \cite{zhuangssr} & Yes & N/A & FPGA & \type{int8} & \type{fp32} & Special Units \\ \midrule
FlightLLM \cite{Zeng2024FlightLLMEL} & Yes & N/A & FPGA & \type{int4} & \type{fp16} & Special Units \\ \midrule
Chen et al. \cite{chenllm} & Yes & N/A & FPGA & \type{int8} & \type{fp16} & \begin{tabular}[c]{@{}c@{}}Special Units\\  (spatial pipeline) \end{tabular} \\ \midrule
\textbf{TATA (Ours)} & \textbf{Yes} & \textbf{No} & \textbf{FPGA} & \textbf{\type{int8}} & \textbf{\type{bfloat16}} & \begin{tabular}[c]{@{}c@{}}\textbf{Reuse \textit{MatMul}}\\  \textbf{Hardware for Vectors} \end{tabular} \\ \bottomrule
\end{tabular}
}
\label{tab:quali-comp-bg}
\end{table}

Various transformer acceleration frameworks for efficient inference have been proposed based on GPU \cite{trans-engine, prabhu2024vattention}, ASIC \cite{hamA3, zadeh2022mokey, Marchisio2023SwiftTron}, and FPGA \cite{wang2022via, zhang2021algorithm, lit2022auto, khannpe, yang2022efa, Zeng2024FlightLLMEL, chenllm, huang2023integer, zhuangssr, swat, marino2023me, hong2022dfx, ltransopu, ftrans, yeetal, flexrun}. Unlike GPU and ASIC design, FPGA has attracted much attention recently, thanks to the configurable and flexible nature of FPGA devices, which has released the low hardware utilization rate issue in fixed architecture GPU or ASIC designs \cite{flexrun}. In terms of hardware architecture, part of existing accelerators focus on linear \textit{MatMul} only, without full support for transformer models \cite{zadeh2022mokey, zhang2021algorithm, lit2022auto, yeetal}. In addition, all other designs with full support for transformer implement individual float-point units \cite{khannpe} or specific modules for non-linear functions \cite{wang2022via, yang2022efa, Marchisio2023SwiftTron, markidis2018nvidia}. Among them, spatial architecture that allows a deep pipeline between different layers has been selected in many previous works \cite{chenllm, zhuangssr, wang2022via}, to reduce off-chip memory I/O. The limited on-chip resources on FPGA challenge such a design choice, especially when the transformer models have a larger and larger scale. On the contrary, \aname utilizes a transferable architecture, allowing full support for all operations in transformer models by compiling non-linear functions into basic operations. Our proposed design reuses integer processing elements for all operations within transformer models, thereby avoiding additional hardware costs for the small-workload non-linear functions, as shown in \figref{fig:bg-trans}. \tabref{tab:quali-comp-bg} presents the qualitative comparison between \aname and the relative works. Note that \aname presents a new but orthogonal angle for transformer accelerators compared to the spatial architecture, and it also has the potential to achieve an efficient pipeline in \aname design.


\section{Motivation}

As shown earlier, while linear layers such as self-attention and MLP can be easily quantized to integers and deployed on matrix multiplication (\textit{MatMul}) processing units, quantizing other non-linear layers without sacrificing model performance is challenging unless one applies quantization-aware training (QAT). Furthermore, maintaining non-linear functions in higher precision (such as floating point) and creating specialized processing units for these less dominating functions results in significant hardware overhead and low hardware utilization. Thus, our primary motivation is: \textit{Can we develop a unified processing unit that efficiently supports linear layers in integer and non-linear layers in floating-point?}

Given a floating-point number $x$ with its significant bit $s_x$, exponent $e_x$ and mantissa $m_x$, the real value of $x$ can be represented as $(-1)^{s_x}\cdot 2^{e_{x}-e_{b}} \cdot m_x$ (the exponent bias is $e_b$). Then, floating-point multiplication (\textit{fpmul}) of two numbers $x$ and $y$ is:

\begin{equation}
    x \cdot y = (-1)^{s_{x} \wedge s_{y}} \cdot 2^{e_{x} + e_{y} - e_{b}} \cdot (m_{x}\cdot m_{y})
    \label{eq:fp-mul}
\end{equation}

In this context, ${e_{x} + e_{y} - e_{b}}$ and $m_{x} \cdot m_{y}$ are operations on integers (specifically, unsigned integers) with a small bitwidth. Consequently, we can implement floating point multiplication using integer processing units with minimal overhead for the significant bit $s$. Standard floating-point addition (\textit{fpadd}), as another basic operation, can be represented as:

\begin{equation}
    \begin{gathered}
    x+y=(-1)^{s_z} \cdot 2^{e_z-e_b} \cdot m_z\\
    e_z=\left\{\begin{array}{ll}
    e_x, & e_x>e_y \\
    e_y, & e_y \geq e_x
    \end{array}, \quad \Delta e= \begin{cases}e_x-e_y, & e_x>e_y \\
    e_y-e_x, & e_y \geq e_x\end{cases} \right. \\
    \left\{s_z, m_z\right\}= \begin{cases}\left\{s_x, m_x\right\}+\left(\left\{s_y, m_y\right\} \gg \Delta e\right), & e_x>e_y \\
    \left\{s_y, m_y\right\}+\left(\left\{s_x, m_x\right\} \gg \Delta e\right), & e_y>e_x\end{cases}
    \end{gathered}
    \label{eq:fp-add}
\end{equation}


In \eqnref{eq:fp-add}, we have already merged the significant bit and mantissa field and transformed this fixed-point number (\type{fxp}) into 2's complement for integer operations. It can be seen that \textit{fpadd} is more complex than \textit{fpmul} due to the alignment of the mantissa. However, after converting to 2's complement, this series of operations becomes integer addition and multiplication. Specifically, the right shift can be performed using \textit{fpmul} with a small lookup table (LUT). The only additional overhead is the conversion between signed digital and 2's complement, as well as the small LUT for right shift.



Since floating-point division is inherently costly, we aim to speed it up using integer operations. To achieve this, we employ the fast inverse square root algorithm \cite{lomont2003fast}, which decomposes division into integer operations, as shown in \eqnref{eq:fpdiv} and Algorithm 1. Observe that the $t^{2}$ computation within this algorithm is distinct from the basic 	\textit{fpmul} and \textit{fpadd} operations. Hence, we designate it as an \textit{approximated} calculation for the square root and division, abbreviated as \textit{fpapp}. This term will be utilized in the subsequent discussion in this paper.


\begin{equation}
\frac{x}{y}=x \cdot \frac{1}{y}=\left\{\begin{array}{c}
x \cdot \frac{1}{\sqrt{y}} \cdot \frac{1}{\sqrt{y}}, \quad y>0 \\
x \cdot \frac{-1}{\sqrt{-y}} \cdot \frac{-1}{\sqrt{-y}}, y<0
\end{array}\right.
\label{eq:fpdiv}
\end{equation}

\begin{algorithm}[tbp]
\small
\renewcommand{\algorithmicrequire}{\textbf{Input:}}
\renewcommand{\algorithmicensure}{\textbf{Output:}}
\caption{Fast Inverse Square Root}\label{alg:isqrt}
\begin{algorithmic}[1]
\Require Input bfloat16 number $y$
\Ensure The inverse square root result $\frac{1}{\sqrt{y}}$
\State $y_{int} = y.view(int16)$ \Comment{Does not change data bits, only changes the data format it refers to}
\State $t_{int} = \texttt{0x5f37} - (y_{int} >> 1)$ \Comment{\texttt{0x5f37} is the magic number in \type{int16} \cite{lomont2003fast}}
\State $t = t_{int}.view(bfloat16)$
\State $\frac{1}{\sqrt{y}} = y  \cdot (1.5 - (y \cdot 0.5 \cdot t^{2})$
\Comment{Define $t^{2}$ as \textit{fpapp} operation. In \aname, \textit{fpapp} is one of the basic operations}
\end{algorithmic}
\end{algorithm}

Based on transformable arithmetic, all basic floating-point operations can be transformed to a series of integer atom operations. As \type{int8} has been the most commonly used format for linear layers quantization, we choose \type{bfloat16} as the high-precision format for non-linear functions, featuring an 8-bit exponent and an 8-bit mantissa. The \type{bfloat16} format has been extensively employed in the deep learning field for many years and has developed into a well-established standard for both training and inference \cite{kalamkar2019studybfloat16deeplearning}. The bitwidth of this unique floating-point format is perfectly aligned with the widely used \type{int8}, for both the exponent and the mantissa. Consequently, based on the analysis aforementioned, it is feasible to repurpose standard \type{int8} processing units for fundamental \type{bfloat16} operations, such as \textit{fpmul}, \textit{fpadd}, and \textit{fpdiv} discussed in this section. We also find that the most commonly used architecture for \textit{MatMul}, systolic array, can actually match the vectorized floating-point execution in terms of computation and data layout. We will further explain the details of hardware design, ISA support, and workload mapping in the following sections.
\section{Hardware Design}

\subsection{System Architecture}
\begin{figure}[tbp]
    \centering
    \includegraphics[width=0.94\linewidth]{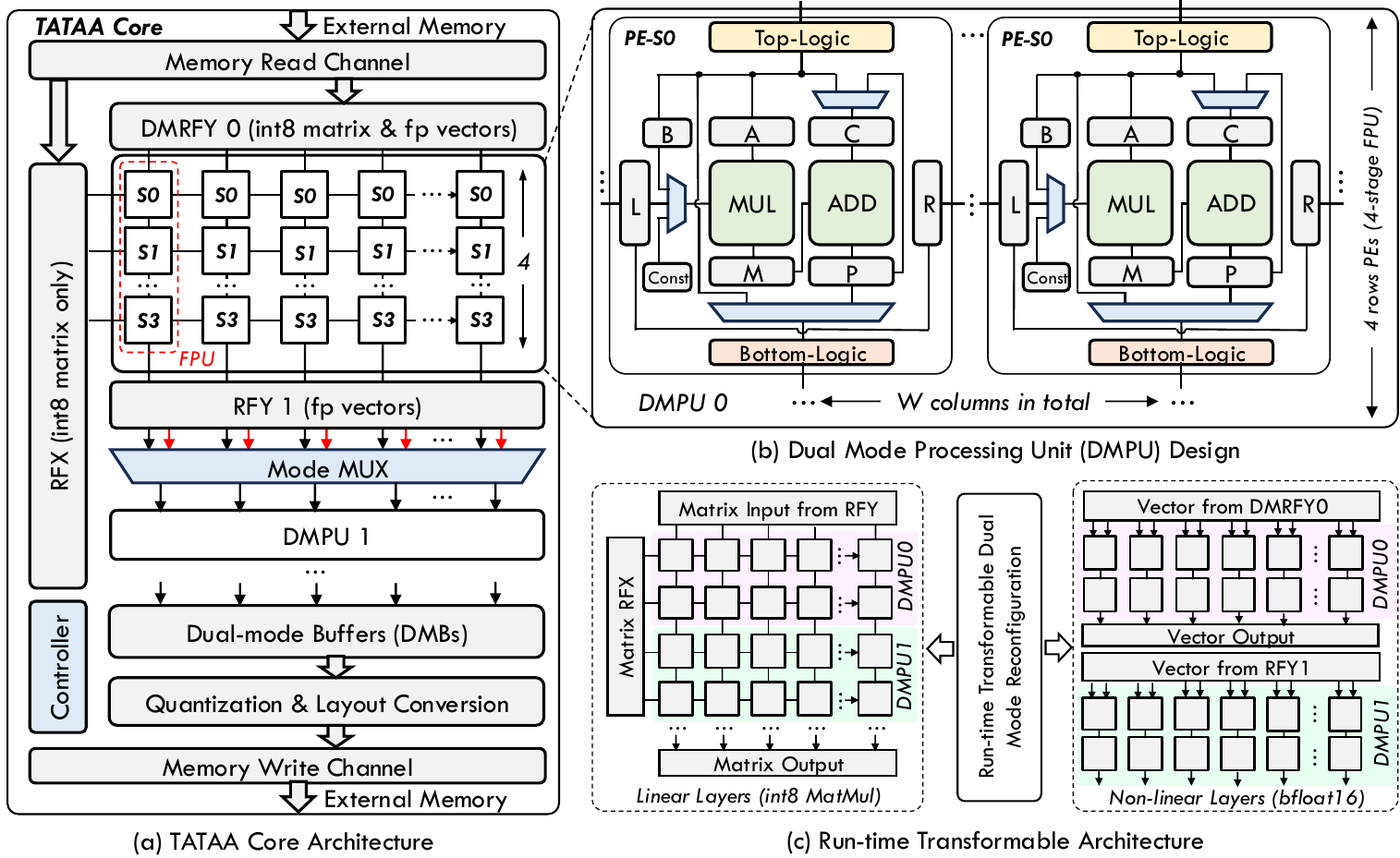}
    \caption{\aname hardware architecture and dual-mode processing unit (DMPU) design. The \aname core can be configured for two kinds of workload during runtime, to support both linear \textit{MatMul} and non-linear functions.}
    \label{fig:sys-arch}
\end{figure}

\figref{fig:sys-arch} illustrates the proposed hardware architecture for the \aname inference scheme. The on-chip accelerator comprises $K$ processing cores that function independently with their own run-time instructions, while only one \aname core is presented for simplification. Given the prevalent use of high-bandwidth memory (HBM) today, we have configured the processing cores with individual memory interfaces to communicate with external memory, thereby optimizing the memory bandwidth utilization rate. To prevent data synchronization issues between cores, we have deliberately divided the workloads among different cores without data dependency, which will be further detailed in our compilation framework. As shown in \figref{fig:sys-arch} (a), each \aname core contains $N$ dual-mode processing units (DMPU) with the integer PE design presented in \figref{fig:sys-arch} (b). The multi-DMPU architecture can be configured for two types of workload in transformer models, as shown in \figref{fig:sys-arch} (c). In \type{int8} \textit{MatMul} mode, all $N$ DMPUs are connected to form a single systolic array. In contrast, in \type{bfloat16} mode, the DMPUs function independently in a SIMD-like manner to execute vectors. The Mode MUX depicted in \figref{fig:sys-arch} (a) manages the run-time configuration that controls the connections between DMPUs with a shared controller. We abstract the on-chip data memory as register files for better high-level abstract and compilation. The input data register files are separated to the X and Y directions (RFX and RFY), corresponding to the horizon and vertical directions in a common systolic array for \textit{MatMul}. They can be configured in different modes and store different formats of data during run-time. Additionally, we incorporate a quantization unit and an on-chip layout conversion module to quantize output results across layers and handle various data layouts between different operations. 

\subsection{Dual-Mode Processing Unit}

The key component in the \aname architecture is DMPU which is configurable for two data formats. As shown in \figref{fig:sys-arch} (b), each DMPU comprises $W$ columns by $4$ rows of the processing element (PE) array. The PE is designed for integer MAC operations, with an integer multiplier (MUL) and a large bitwidth adder (ADD), as standard setups in \type{int8} \textit{MatMul}. In the \type{int8} \textit{MatMul} mode, all the $N$ arrays in the DMPUs are connected to function as a unified $W$ by $4N$ systolic array, and the results of the bottom PEs in one DMPU will be fed to the next DMPU as the top input, controlled by the mode MUX in \figref{fig:sys-arch} (a). Based on the conventional integer MAC PE design, we expand it by adding low-overhead top and bottom logic units and some extra MUX to support mapping \type{bfloat16} operations into it. To enhance the throughput of \textit{MatMul} and address the memory-bound challenge in transformer models, we have implemented two loading ports per \aname core to interact with external memory, given the presence of two matrices in \textit{MatMul}. It's important to highlight that these memory ports are not fixed to specific register files, RFX or RFY; instead, they are dynamically managed by a crossbar which routes the register files to the appropriate port. 

When \aname works in the \type{bfloat16} mode, each DMPU functions autonomously following a SIMD-like process, receiving data from the corresponding RFY instead of the previous DMPU. Each column of the PE array is considered a floating point unit (FPU), and the $4$ rows become $4$ pipeline stages in a floating point, naming PE-S0 (the first stage), PE-S1, PE-S2, and PE-S3 (the last stage). The results are buffered in dual-mode buffers (DMB) in both modes before being stored back to external memory. Note that the intermediate \type{bfloat16} results can also be written to RFY for further computation, avoiding frequent I/O access. Consequently, the core utilizes $W \cdot N$ parallel SIMD lanes in the \type{bfloat16} mode, allowing the software stack to specify vectorized operations in \type{bfloat16} with a maximum vector length of $W \cdot N$. The switch of execution modes is completely online without reconfiguring the hardware, as presented in \figref{fig:sys-arch} (c), thanks to the custom ISA design in \aname.

\subsection{Processing Element Design on FPGA}

\begin{figure}[tbp]
    \centering
    \includegraphics[width=0.85\linewidth]{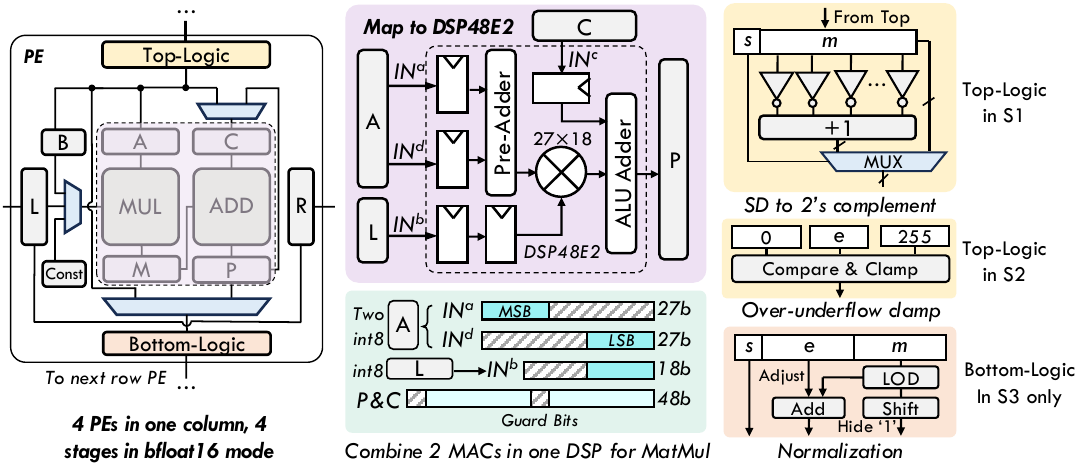}
    \caption{Processing element (PE) design in the proposed DMPUs when deployed on FPGA devices. The converter from 2's complement to signed digit (SD) logic that locates in PE-S3's bottom logic is not depicted in this figure since it is similar to the SD to 2's complement circuit in PE-S1.}
    \label{fig:pe-design}
\end{figure}

\figref{fig:pe-design} shows the central component of DMPU in \aname, specifically, the processing element (PE). Within this PE, the multiplier (MUL) and adder (ADD) perform multiply-accumulation (MAC) tasks for \textit{MatMul} layers, alongside executing basic integer multiplication or addition in the \type{bfloat16} mode according to the breakdown described from \eqnref{eq:fp-mul} to \eqnref{eq:fpdiv}. As \aname aims to reuse the same hardware units for these diverse functions, multiple MUX are incorporated within the PE to manage data pathways in different modes. Additionally, top- and bottom-logic are executed through look-up tables (LUTs) on FPGA for additional operations with minimal overhead, such as normalization and overflow or underflow clamping, vital for all floating-point calculations. Moreover, since DSP48E2 block in modern AMD FPGAs has large multiplier \& adder bitwidth ($27\times 18$ for multiplier and $48$-bit adder), a combined MAC optimization is implemented in each PE to enhance runtime throughput, a strategy commonly adopted in various FPGA accelerators \cite{int8opt, hikonv}. When functioning in \type{MatMul} mode, the PEs are organized as a large systolic array, integrating multiple DMPUs to maximize throughput.

As demonstrated previously, \type{bfloat16} operations are perceived as a four-stage pipeline, with each stage assigned to one processing element (PE) within a column. To utilize the same hardware, operations in \type{bfloat16} format must be transformed from the original signed-digit format of the floating-point standard into the 2's complement format used in DSP blocks, with conversions back to signed digits required before storing the results in memory. Consequently, the PE must include additional processing units specifically for \type{bfloat16} mode, identified as top-logic and bottom-logic in \figref{fig:pe-design}. It is important to highlight that the top-logic architecture varies across different PE stages, and the extra circuits have minimal overhead. For example, the converter from signed digit (SD) to 2's complement is implemented in PE-S1 (second stage), while the logic handling overflow and underflow clamping is placed in PE-S2, and only PE-S3 contains the bottom logic needed for final normalization prior to outputting the \type{bfloat16} result. In detail, the SD to 2's complement converter in \figref{fig:pe-design} concludes with a bitwise inverter, a $+1$ adder, and a MUX to select positive or negative data as the output. In addition, since the exponent in \type{bfloat16} is 8-bit, the corresponding PE needs to clamp exponent from $0$ to $255$, thus implementing such a unit in top-logic. The overflow \& underflow The normalization unit is the same as the standard normalization design in common floating-point units, with a leading one detector to align the mantissa and hide the hidden `1' and an adder to adjust exponent after shifting mantissa. In addition, each PE contains several constant registers for \type{bfloat16} mode in each stage, as the input of multiplier. The detailed data flow of \textit{MatMul} mode and \type{bfloat16} mode will be thoroughly discussed in the following sections, explaining how these top- and bottom-logic places in different stages of PEs.

\begin{figure}[tbp]
    \centering
    \includegraphics[width=0.76\linewidth]{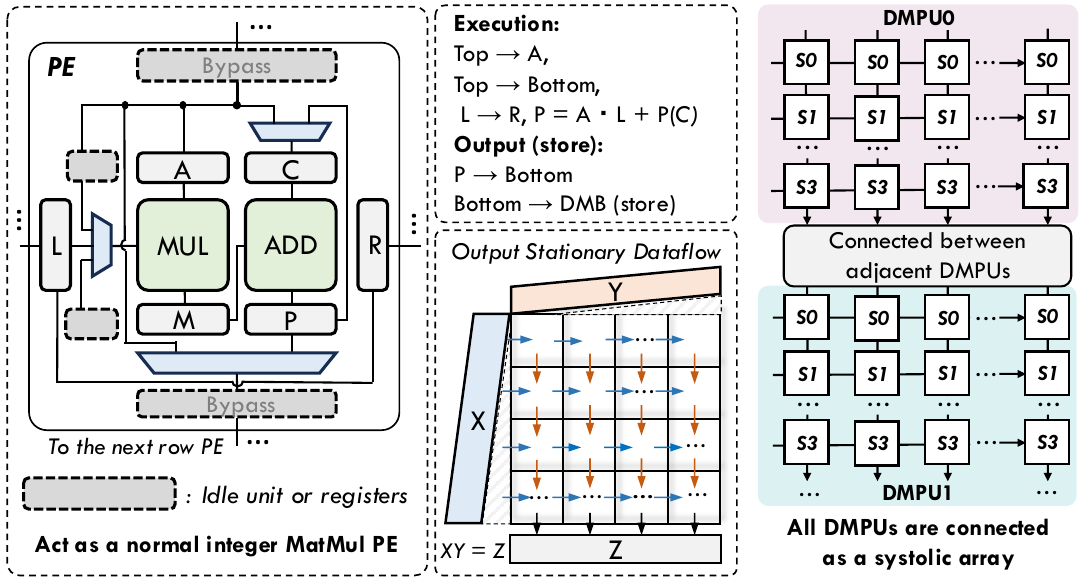}
    \caption{Execution dataflow in \type{int8} \textit{MatMul} mode, in which all the PEs are connected as a whole systolic array and deploy output stationary dataflow.}
    \label{fig:int8-flow}
\end{figure}

\subsection{Dataflow}

The architecture proposed in \aname can be configured for \type{int8} \textit{MatMul} mode and \type{bfloat16} mode during runtime. \figref{fig:int8-flow} shows the dataflow in the \type{int8} \textit{MatMul} mode, where all PEs are connected as a systolic array, and intermediate results accumulate across DMPUs. We choose to deploy the output stationary dataflow for matrix multiplication. In such an execution flow, the $\mathbf{X}$ and $\mathbf{Y}$ matrices go through the systolic array in the X (horizontal) and Y (vertical) directions, respectively. Registers L and R are responsible for horizontal and vertical data passing, while the bottom register directly accepts the data from the top and sends them to the next PE. After \textit{MatMul} finishes, the results stored in register P will be sent to the corresponding dual-mode buffer (DMB). The intermediate sums are accumulated in the \type{int16} format and subsequently quantized to either \type{int8} or \type{bfloat16} before being saved to external memory, depending on the format required by the subsequent layer. The static scaling factors are pre-loaded to the quantization unit in the \aname core before \textit{MatMul} starts.

\begin{figure}
    \centering
    \includegraphics[width=0.65\linewidth]{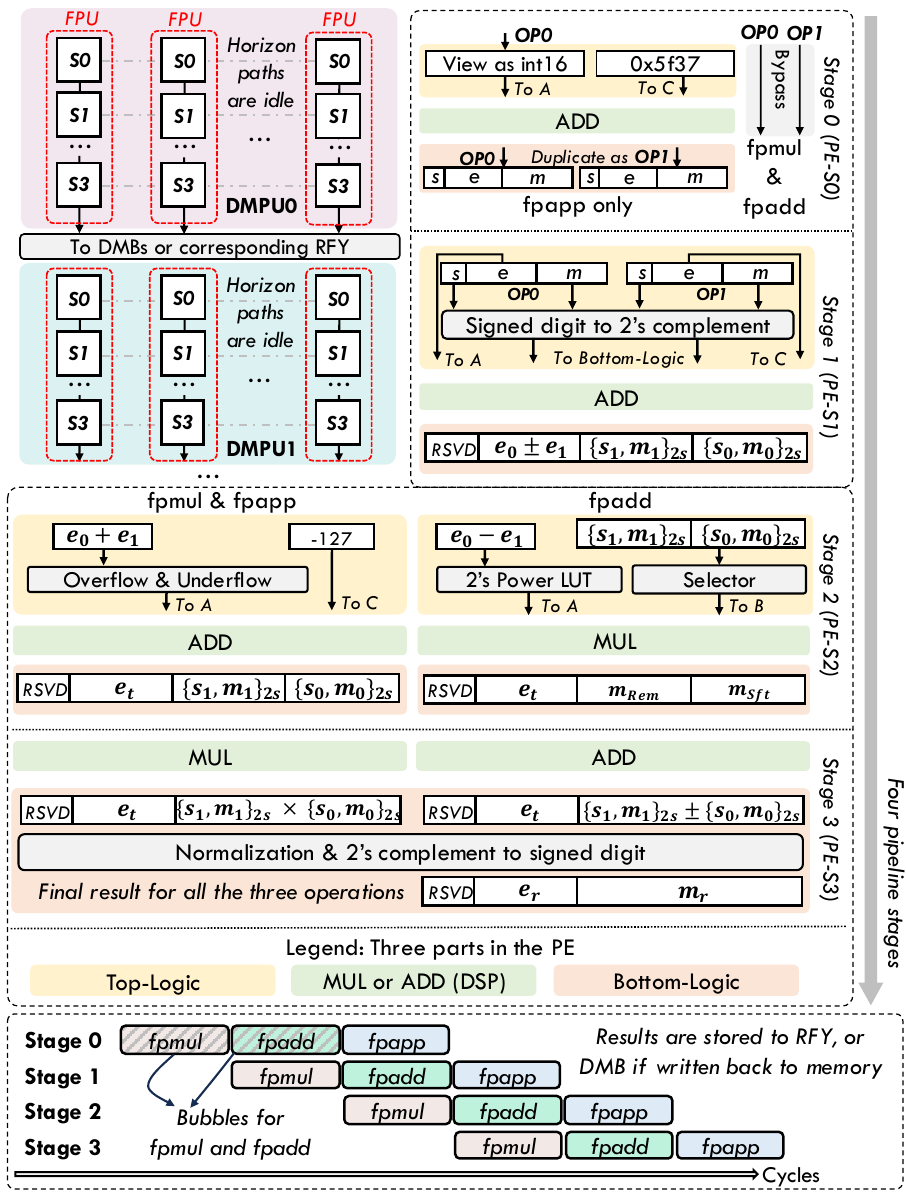}
    \caption{\type{bfloat16} mode dataflow and how to reuse the integer processing units. The Top-Logic, MUL \& ADD, and Bottom-Logic processing are depicted in specific colors in this figure.}
    \label{fig:fp-flow}
\end{figure}

When the \aname architecture is set in \type{bfloat16} mode, it can execute three basic operations: multiplication (\textit{fpmul}), addition (\textit{fpadd}) and the approximation step for the inverse square root in \eqnref{eq:fpdiv} and Algorithm 1 to support $(0x5f37-(y_{int} >> 1))^{2}$ (\textit{fpapp}) operation. These operations can be assigned directly to the $4$ pipeline stages in the $4$ rows of integer-based PE, as illustrated in \figref{fig:fp-flow}. Thanks to the arithmetic analysis from \eqnref{eq:fp-mul} to \eqnref{eq:fpdiv}, we can convert \type{bfloat16} operations into a sequence of integer operations. The integer multiplier and adder (MUL \& ADD in \figref{fig:fp-flow}) are reused in the \type{bfloat16} mode for higher resource efficiency. The floating-point pipeline not only adapts to integer arithmetic but also shares similar extra top- and bottom-logic, significantly reducing hardware overhead. The overall overhead encompasses converters for signed digits and 2's complement, a compact 2's power LUT, typical overflow and underflow management, and a normalization unit, all of which are standard components in conventional floating-point units. Specifically, the special \textit{fpapp} first treats the input \type{bfloat16} binary number as \type{int16}, performs integer subtraction (addition) in the first stage, and converts the integer binary number back to \type{bfloat16} with the remaining $3$ stages for square. Using this arithmetic mapping, the proposed DMPU can execute one operation per column in parallel, thus improving the SIMD execution of \type{bfloat16} operations from a global perspective. This type of sharing scheme between two modes significantly reduces the consumption of hardware resources so that \aname can map more parallel cores when resources are limited. 

\subsection{Register Files and Buffers}

\begin{figure}[tbp]
    \centering
    \includegraphics[width=0.82\linewidth]{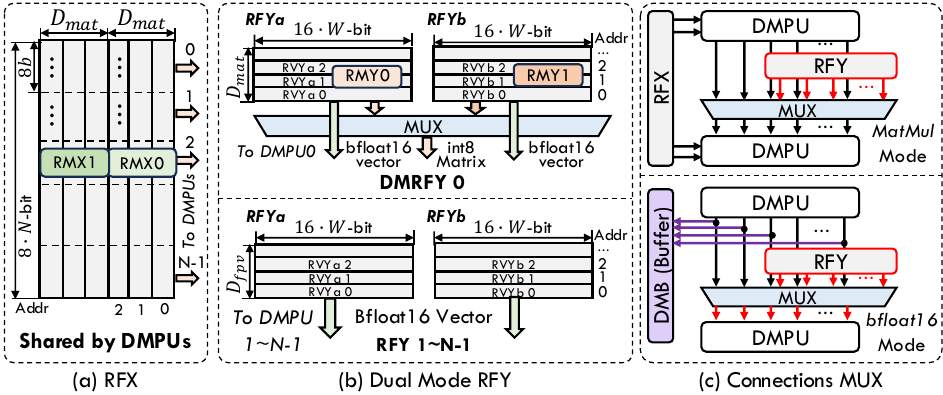}
    \caption{Register files (RF) design in \aname, and the connections between DMPU, RF and Dual-mode Buffers in different modes.}
    \label{fig:dmrf-diag}
\end{figure}


Since the proposed \aname framework is for flexible acceleration, abstract register files are required for efficient instruction set architecture (ISA) and compiler design. As shown in \figref{fig:dmrf-diag} (a), the RFX is available only in the \type{int8} \textit{MatMul} mode since the \type{bfloat16} data only go through the Y direction. Therefore, the abstract concept \textit{"registers"} in RFX is actually matrix buffers, and we set two registers inside (RMX0 and RMX1) to apply double-buffer optimization, hiding the memory I/O latency. The RFX has $N$ ports to send data to corresponding $N$ DMPUs. In \aname, we define that each of the X matrix buffers has $D_{mat}$ depth for matrix multiplication, so the total depth of RFX is $2\cdot D_{mat}$.


In terms of RFY, we set up two memory banks named RFYa and RFYb, to support the two input operators in the \type{bfloat16} mode, and each address stores one part of the parallel vector, as shown in \figref{fig:dmrf-diag} (b). The data layout becomes more complex because only the RFY for DMPU0 (Dual-mode RFY, DMRFY0) needs to store both \type{int8} matrices and \type{bfloat16} vectors. When it works in \type{int8} \textit{MatMul} mode, the DMPU only needs one specific part of RFYa and RFYb. The two-bank design naturally supports double-buffer optimization (selected by the MUX), so each of them only costs $D_{mat}$ depth. Like RFX, we abstract the matrix buffers as RMY0 and RMY1 physically corresponding to RFYa and RFYb. For the other RFY, they can only be used in the \type{bfloat16} mode. Hence, the depth of these register files is set to $D_{fpv}$ ($D_{fpv} \ll D_{mat}$), so the extra memory overhead of these \type{bfloat16} vectors is relatively small. There is a natural data layout conflict between \type{int8} and \type{bfloat16}. Since each \type{bfloat16} number takes 16 bits, the output of one DMRFY0 bank (RFYa or RFYb) should have $W \cdot 16$ bits in \type{bfloat16} mode. However, DMRFY0 also needs to store the \type{int8} matrix, and the $W$ columns of the PE array only need $W \cdot 8$ bits. Thanks to the combined MAC optimization introduced before, the bitwidth of the DMPU input in \type{int8} \textit{MatMul} becomes $W \cdot 16$-bit, matching the bitwidth of one bank in DMRFY0. As for non-FPGA implementation, designers can also deploy such optimization with larger multiplier and accumulator, to fit bitwidth in \type{bfloat16}, as well as benefits from higher throughput in \textit{MatMul} operations.

In \aname architecture, the dual-mode buffers (DMB) serve the function of storing results temporarily before they are returned to the external memory. Each DMPU has a corresponding DMB in the bottom output direction. Importantly, DMBs have varying execution procedures in \textit{MatMul} and \type{bfloat16} modes. In \textit{MatMul} mode, all DMPUs form a single systolic array, leading to inactivity in the DMBs linked to DMPU $0$ through DMPU $6$, with only the last DMPU $7$ receiving \textit{MatMul} intermediate results. In contrast, in \type{bfloat16} mode, all DMPUs with $W$ columns (essentially, $W$ FPUs) function independently following a SIMD approach, necessitating all DMBs to store \type{bfloat16} vector results. A MUX also dictates the data path between the two modes, as illustrated in \figref{fig:dmrf-diag} (c). In \textit{MatMul} mode, both the input \textbf{Y} and output \textbf{Z} traverse all DMPUs, with the bottom DMPU receiving data transmitted from the top DMPU as determined by the MUX. Conversely, in \type{bfloat16} mode, each DMPU obtains its input from RFY selected by the MUX.

\subsection{Layout Conversion and On-chip Quantization}

\begin{figure}[tbp]
    \centering
    \includegraphics[width=0.62\linewidth]{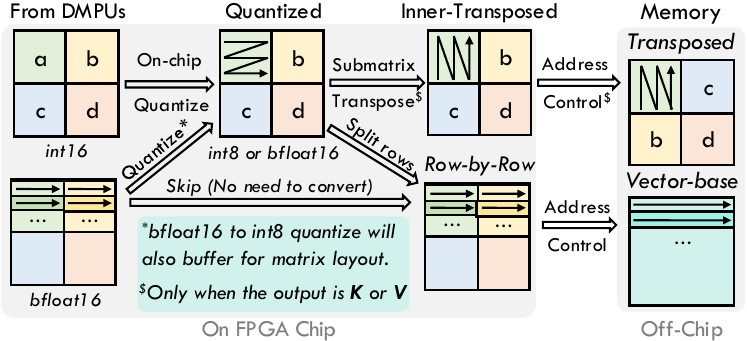}
    \caption{Illustration of the on-chip quantization and layout conversion module.}
    \label{fig:quant-lc}
\end{figure}

Before writing the calculated \type{int8} matrix or \type{bfloat16} vector results back to external memory, the quantization unit dynamically quantizes the activations according to the current configuration, as shown in \figref{fig:quant-lc}. \aname architecture supports data format switching between \type{int8} and \type{bfloat16}, with four types of configuration in the quantization unit. All the conversions here can be handled based on \eqnref{eq:quant} with preloaded floating point scaling factors, except the \type{bfloat16} $\rightarrow$ \type{bfloat16} conversion does not require any specific quantization. Besides, if the subsequent workload cannot be deployed directly on the current data layout, \aname handles on-chip layout conversion. On the one hand, the QK-MUL layer requires the matrix \textbf{K} from the previous layer to be transposed to match the layout. On the other hand, the \type{bfloat16} mode is aimed at a vector-based layout, which differs from matrix workloads, which requires hardware with a row-by-row storage scheme for both \type{int8} $\rightarrow$ \type{bfloat16} and \type{bfloat16} $\rightarrow$ \type{int8} conversion. The only hardware overhead is to transpose the submatrix from DMPU, since all other conversions can be done by delicately controlling the write-back addresses.
\section{Compilation}


\begin{table}[tbp]
\centering
\caption{\aname ISA}
\resizebox{0.57\linewidth}{!}{
\begin{tabular}{cc}
\toprule
Instruction Type & Description \\ \midrule
CONFIG & Set up static parameters (e.g., scaling factors, constants) \\ \midrule
LOAD.M & Load a matrix from memory \\ \midrule
LOAD.V & Load a vector from memory \\ \midrule
MATMUL & Execute matrix multiplication $\mathbf{Z}=\mathbf{XY}$ \\ \midrule
MUL.V & Execute \textit{fpmul} of two vectors \\ \midrule
ADD.V & Execute \textit{fpadd} of two vectors \\ \midrule
APP.V & Execute \textit{fpapp} of one vector \\ \midrule
STORE.M & Store a matrix (executed results) to memory \\ \midrule
STORE.V & Store a vector (executed results) to memory \\ \bottomrule
\end{tabular}
}
\label{tab-isa}
\end{table}

Before introducing the proposed \aname compilation framework, we define the terminology related to layers, nodes, and operations. A layer is a concept at the model level, with its definition detailed in \figref{fig:bg-trans}. The nodes operate at the graph level and are derived from a specific transformer model. For example, the non-linear SoftMax function can be broken down into a sequence of sub-functions, such as exponentiation, summation, and division, that become nodes in the computational graph. These nodes can be amalgamated or subdivided into additional nodes. In matrix multiplication (\textit{MatMul}), a node with a large \textit{MatMul} size can be divided into smaller tiled \textit{MatMul} to better align with hardware structures. Operations reflect a hardware-level concept derived from nodes, implemented in the \aname architecture, as discussed in Section 4.

\subsection{Instruction Set Architecture (ISA)}

To better decouple hardware and software, we have developed a customized Instruction Set Architecture (ISA). Our software system can map linear layers in \type{int8} and non-linear operations with a high-precision approximation in \type{bfloat16}. \tabref{tab-isa} presents the simple ISA design in \aname. In an ISA-level perspective, the controller is able to detect data dependencies and exploit instruction-level parallelism (ILP) to improve throughput performance. As an example, the double buffer optimization allows the parallel execution of the LOAD.M and MATMUL instructions. Furthermore, the previously mentioned data layout conversion with specific write-back addresses is incorporated into the STORE.M and STORE.V instructions, offering sufficient flexibility and comprehensive support for inference runtime. After compilation, all the runtime instructions are stored in the external memory, and \aname accelerator acts as a processor that fetches instructions from external memory, removing the requirement overhead of host-based scheduling.

\subsection{End-to-End Transformer Mapping}

\begin{figure}[tbp]
        \centering
        \includegraphics[width=0.60\linewidth]{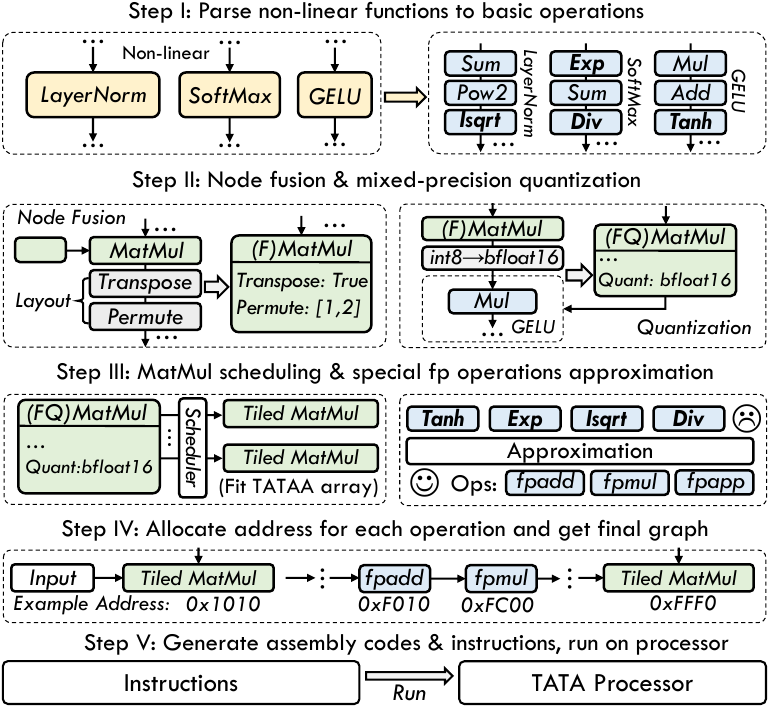}
        \caption{Top-Down workflow of \aname compiler. Note that \aname supports various non-linear functions. The depicted LayerNorm, SoftMax and GELU are used as examples.}
        \label{fig:sys-compiler}
\end{figure}

\figref{fig:sys-compiler} illustrates the top-down compilation process from an input transformer-based model to the \aname hardware runtime. The compilation framework first parses and converts nonlinear functions into a series of basic operations (e.g., summation, squaring, multiplication, etc.) by examining the computation graph. For example, the LayerNorm function in \eqnref{eq:non-linear} is parsed to summation (sum up vectors in-between), power of 2 (calculate $\mathbf{x}^{2}$ for variantion), division (calculate $E[\mathbf{x}]$ based on summation), etc., as a series of operations. This parsing process has been mature in existing machine learning frameworks like ONNX \cite{bai2019}.

\begin{algorithm}[tbp]
\small
\renewcommand{\algorithmicrequire}{\textbf{Input:}}
\renewcommand{\algorithmicensure}{\textbf{Output:}}
\caption{Approximation examples for non-linear functions}\label{alg:nonlinear}
\begin{algorithmic}[1]
    \Require Input activation $x$
    \Ensure Exponent value of $x$, $exp\_x$
    \State $exp\_x = 2^{\floor{(x / \ln{2})}}$ \Comment{$2^{\floor{\cdot}}$ is fused into output quantization process by a small LUT}
    \Ensure Inverse square root of $x$, $isqrt\_x$
    \State $y = \texttt{0x5f37} - (short(x) >> 1)$ \Comment{Similar to \textbf{Algorithm 1}}
    \State $isqrt\_x = 1.5y - 0.5x \cdot y^{3}$
    \Ensure Padé approximation of $\tanh{(x)}$
    \State $tanh\_x = clamp(\frac{27x + x^3}{27 + 9x^2}, min=-1, max=1)$ 
\end{algorithmic}
\end{algorithm}

Next, the compiler applies node fusion and mixed-precision quantization by integrating data layout conversion and quantization into the previous node, because the hardware supports on-chip quantization \& layout conversion at runtime. With this intermediate representation (IR), the compiler then schedules linear \textit{MatMul} operations into a sequence of tiled \textit{MatMul} operations, with each tile conforming to the size of the \aname systolic array. Currently, the basic operations of nonlinear functions are approximated and compiled into \aname-supported operations (i.e., \textit{fpmul}, \textit{fpadd}, \textit{fpapp}). Details of how to approximate these operations are shown in Algorithm \textbf{2}. Upon completing these conversions, the compiler can analyze \type{bfloat16} workloads and vectorize them for SIMD-like instructions MUL.V and ADD.V as shown in \tabref{tab-isa}. Finally, the compiler assigns addresses to each atomic operation for the hardware runtime and generates binary instructions for the \aname processor.


\subsubsection{MatMul Schedule}

\begin{figure}[tbp]
    \centering
    \includegraphics[width=0.6\linewidth]{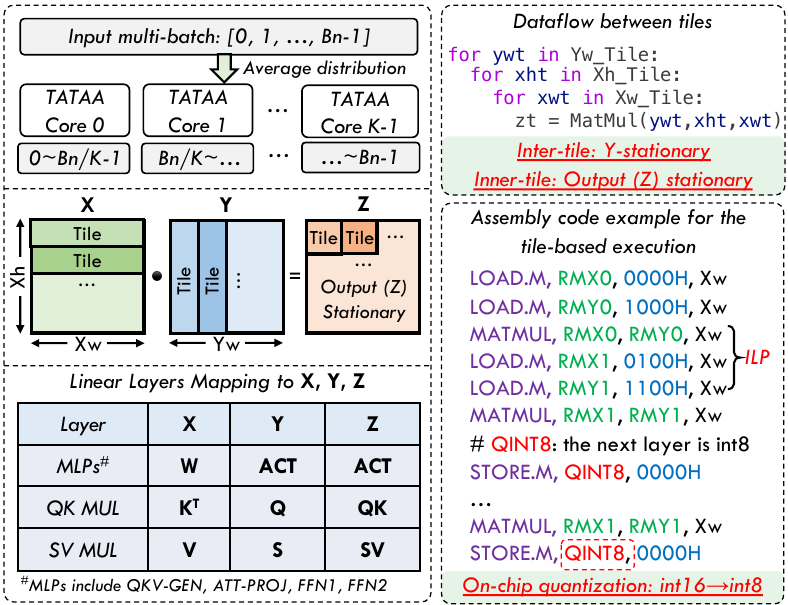}
    \caption{Linear \textit{MatMul} scheduling in \aname compilation.}
    \label{fig:compiler-opt}
\end{figure}

The scheduler first analyzes the shape of the output matrix and distributes it evenly across batches to avoid data dependency between parallel cores. As activations are independent across various batches, the \textit{MatMul} scheduler and the non-linear functions compiler can concentrate solely on the batches of a single core, incrementally updating the starting address to determine the activation addresses for other cores. Each \aname core employs \textit{MatMul} based on an output-stationary dataflow with a fixed output tile size, $W$ by $4N$. Based on this tiling, each output tile corresponds to a tile of matrices \textbf{X} and \textbf{Y}. After determining the addresses for \textbf{X}, \textbf{Y}, and \textbf{Z}, the scheduler generates a series of instructions LOAD.M, MATMUL, and STORE.M as assembly codes for runtime inference and applies double buffer optimization by reordering the three types of instruction, similar to the instruction-level parallelism (ILP) strategy. \figref{fig:compiler-opt} illustrates the scheduler in terms of dataflow design and an example of ILP optimization using assambly codes. In this example, the \texttt{MATMUL,RMX0,RMY0,Xw}, \texttt{LOAD.M,RMX1,0100H,Xw} and \texttt{LOAD.M,RMY1M,1100H,Xw} can be executed in parallel since 1) there is no register files index conflict 2) we design two I/O ports for loading. Besides, the scheduler needs to decide how to map the two operands of \textit{MatMul} into \textbf{X} and \textbf{Y}, since the data layout of the two input ports in systolic array are different. For normal MLP, we map the weights and activations into \textbf{X} and \textbf{Y}, respectively, while for QK-MUL and SV-MUL without weights, we use another mapping scheme, as the layout conversion between \textbf{X} and \textbf{Y} can be done on hardware. 

\subsubsection{Non-Linear Functions Optimization}

\begin{figure}[tbp]
    \centering
    \includegraphics[width=0.78\linewidth]{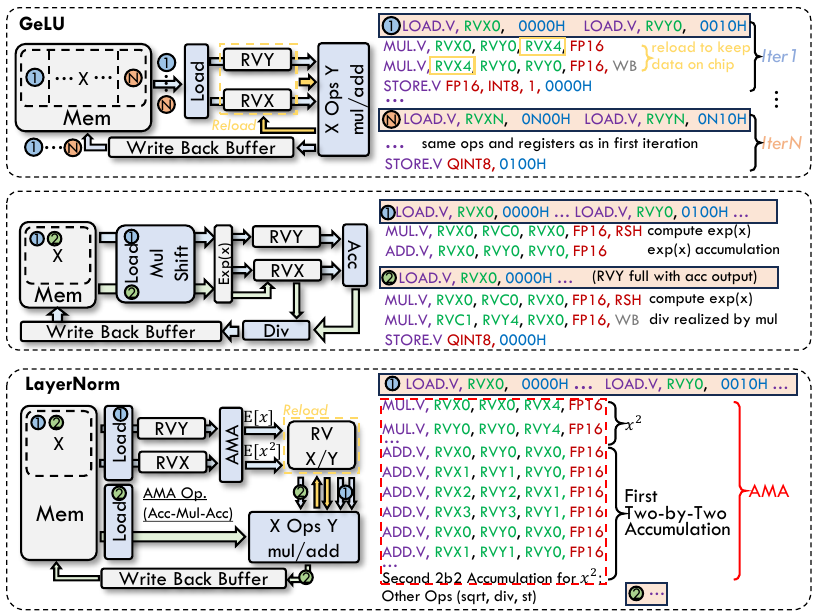}
    \caption{Non-linear functions compilation in \aname. We only present three examples which are commonly applied in most transformer models.}
    \label{fig:compiler-nonl}
\end{figure}

The fundamental process of mapping non-linear functions to \aname hardware runtime is illustrated in \figref{fig:compiler-nonl}. To optimize the performance of executing non-linear functions, TATAA focuses on reducing memory access and maintaining computation on the chip by consistently reloading computation results into local registers. As shown in the yellow dashed box with the reload operation in \figref{fig:compiler-nonl}, when two vectors perform a MUL.V or ADD.V operation, the result is stored (reloaded) in one of the X or Y registers. By consistently performing this reloading and computation process, most computational operations are executed together without requiring additional memory access. All operations that do not involve memory access employ this on-chip computation method to maximize execution throughput. Since this on-chip computation necessitates loading as much data as possible into the register files before executing computation instructions, load instructions are assembled at the beginning of each non-linear function. To mitigate the sluggish memory access times for these load instructions, the outstanding transaction features of AXI are utilized to minimize the total load duration.

Additionally, we emphasize two types of node-level optimization for non-linear functions, as depicted in \figref{fig:compiler-nonl}, with the aim of significantly reducing memory I/O costs while maximizing computational efficiency. We optimize the compiler to reuse input \textbf{x} vectors in register files by rearranging the computation nodes within the graph. This method prevents redundant loading of \textbf{x} onto the hardware. Given that the LayerNorm function involves both variance and mean calculations, as shown in \figref{fig:compiler-nonl}, AMA (accumulate-multiply-accumulate) facilitates the computation of $E[\mathbf{x}]$, $\mathbf{x}^{2}$, and $E[\mathbf{x}^{2}]$ by splitting both the X and Y registers into two groups. The input vector \textbf{x} is first loaded into the first group of the X and Y registers and then multiplied by itself to store the result $\mathbf{x}^2$ into the second group of registers. As demonstrated in the assembly code, after loading the input vectors, the AMA process initiates with a series of MUL.V instructions to compute $x^2$. The initial and subsequent two-by-two accumulation operations are executed without encountering data hazards, thus avoiding additional delays between computation instructions. In the first accumulation example, the values in the X registers 0 to 3 are added to the corresponding values in the Y registers. The sums are then stored back into registers 0 to 1 of both X and Y registers in a crosswise manner. This approach effectively reduces the total number of registers holding partial sums by half. The values in registers 0 to 1 are further consolidated into another partial sum, which is reloaded into register 0 in both X and Y registers. This partial sum is subsequently added together to produce the final accumulation result. Since the values from the first and second accumulation processes are stored in separate sections of the X and Y register files, the second two-by-two accumulation can occur concurrently with the first accumulation process. This approach ensures that the maximal amount of input vector data that both register files can hold is loaded only once from memory, leaving a substantial portion of the computation to be performed on-chip, thereby maximizing performance. Furthermore, to mitigate potential data hazards during this consistent computation, a two-by-two accumulation method is employed, which adds every two lines of input vectors and stores the results crosswise into the X and Y registers. Such an optimization compilation works for other normalization functions, e.g., RMSNorm as well.

Furthermore, it is important to note that the GELU function maintains a consistent tensor shape across all nodes, enabling segmentation of all nodes into uniform tile shapes and allowing sequential execution of these tiles from start to finish. As illustrated in \figref{fig:compiler-nonl}, each iteration corresponds to a tile, and within every iteration, a much longer sequence of computation instructions is executed between a load and a store instruction. The yellow arrows indicate this extended sequence of computations. Although the number of memory accesses increases with the number of tiles, the memory overhead remains minimal compared to the lengthy computation sequence. Upon completion of each iteration, the results are written back to the address from which the input vector was initially read. Simultaneously, all registers used in the current iteration are cleared and ready to receive the next tile for the subsequent iteration. This multi-iterative method is optimized for GELU scheduling to achieve two main objectives: 1) avoiding intermediate I/O communication with external memory and 2) accommodating the limited register file space. Given that \aname employs a layer-by-layer execution method, these I/O optimizations are crucial for improving throughput efficiency. Note that as long as the activation functions have the same tensor shape patterns across all nodes (e.g., SiLU), this tile-based compilation can be applied.

\section{Evaluation}

\subsection{Experiments Setup}

\begin{table}[tbp]
\centering
\caption{Selected Transformer Models or Non-linear Functions in the Experiments}
\label{tab:model-dim}
\resizebox{0.93\linewidth}{!}{%
\begin{threeparttable}
\begin{tabular}{@{}ccccccc@{}}
\toprule
Model & Type & \# Blocks & \# Heads & Hidden Size & MLP Size & Non-linear Functions \\ \midrule
Deit-S & Encoder & 12 & 6 & 384 & 1536 & \multirow{5}{*}{\begin{tabular}[c]{@{}c@{}}SoftMax \\ LayerNorm \\ GELU \end{tabular}} \\
Deit-B & Encoder & 12 & 12 & 768 & 3072 & \\
Swin-T & Encoder & $^{\dagger}$\{2,2,6,2\} & $^{\dagger}$\{3,6,12,24\} & $^{\dagger}$\{96,192,384,768\} & $^{\dagger}$\{$56^2$, $28^2$,$14^2$,$7^2$\} & \\
BERT & Encoder & 12 & 12 & 768 & 3072 &  \\
GPT2 & Decoder & 24 & 16 & 1024 & 4096 &  \\ \midrule
OPT-1.3B$^{\#}$ & Decoder & 24 & 16 & 2048 & 8192 & SoftMax, LayerNorm, ReLU \\ \midrule
Llama-7B$^{\#}$ & Decoder & 32 & 32 & 4096 & 11008 & SoftMax, RMSNorm, SwiGLU \\ \midrule
ChatGLM2$^{\#}$ & Decoder & 28 & 32 & 4096 & 13696 & SoftMax, RMSNorm, SiLU \\ \bottomrule
\end{tabular}%
\begin{tablenotes}
\footnotesize
\item[$\dagger$] \{\dots\} shows dimension variance of each stage in a Swin-T \cite{liu2021swin}.
\item[$^{\#}$] These large language models are not evaluated on hardware runtime. We only select them for various non-linear functions test.
\end{tablenotes}
\end{threeparttable}
}
\end{table}

We implemented and prototyped \aname on Alveo U280 FPGA platform using Verilog HDL and Vitis 2021.1 tools under $225$ MHz frequency, to measure resource utilization, power consumption, and end-to-end runtime throughput. The hyperparametes mentioned in \figref{fig:sys-arch} are set as $K=8$, $N=8$, $W=16$, with a corresponding $32$ by $32$ systolic array and $128$-lane SIMD FPUs in each core. In Alveo U280, we set up $16$ AXI channels for the $8$ \aname cores and each AXI channel has $256$-bit memory bitwidth.

We have chosen a range of transformer models to evaluate the accuracy of quantization and their runtime performance in tasks such as image classification, text classification, and text generation. The selected models are listed below and their feature dimensions are shown in \tabref{tab:model-dim}.


\begin{itemize}
    \item For Vision Transformers (ViT), we select DeiT\cite{touvron2021training} and Swin Transformer (Swin) \cite{liu2021swin} with ImageNet-1k \cite{deng2009imagenet} dataset for the image classification.
    \item BERT \cite{devlin2018bert}, as a widely used language model, is also selected for evaluation based on the GLUE benchmark including different tasks \cite{wang2018glue}.
    \item We also evaluate other popular language models, GPT-2 \cite{radford2019language} and OPT \cite{zhang2022opt}, for the text generation task with LAMBADA \cite{paperno2016lambada} and WikiText-2 datasets.
    \item To evaluate the general support of \aname for non-linear functions in transformer models, we also select state-of-the-art Llama \cite{touvron2023llama} and ChatGLM2 \cite{glm2024chatglmfamilylargelanguage} where some extra functions like RMSNorm \cite{zhang-sennrich-neurips19}, SwiGLU \cite{shazeer2020glu} and SiLU \cite{ramachandran2017searching} are deployed. Note that the two models are not evaluated end-to-end, as we only tested the non-linear functions part on hardware.
\end{itemize}

\begin{figure}[tbp]
    \centering
    \includegraphics[width=0.9\linewidth]{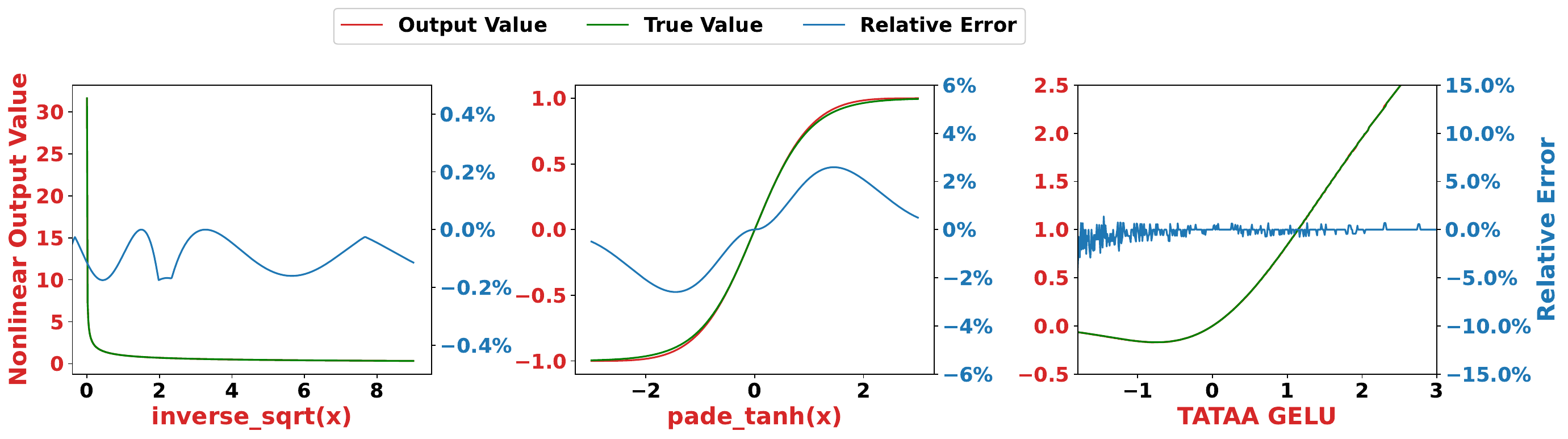}
    \caption{The nonlinear function precision measures between our approximated and PyTorch's built-in functions. We selected two approximated sub-operations (inverse square root and pade $tanh$), and a function-level GELU function for error evaluation.}
    \label{fig:nonlinear_funtion_error}
\end{figure}

\subsection{Model Accuracy}

\begin{table}[tbp]
\caption{Quantization Evaluation for Various Transformer Models Based on \aname Setups with \type{int8} + \type{bfloat16}}
\resizebox{\linewidth}{!}{
\begin{tabular}{@{}cc|cccccc|ccc|c|cc@{}}
\toprule
\multirow{2}{*}{Method} & \multirow{2}{*}{\begin{tabular}[c]{@{}c@{}}PTQ.\\ Format\end{tabular}} & \multicolumn{6}{c|}{ViT Classification Accuracy (\%)} & \multicolumn{3}{c|}{BERT on GLUE (\%)} & GPT-2 Medium & \multicolumn{2}{c}{OPT-1.3B$^{\dagger}$} \\ \cmidrule(l){3-14} 
 &  & DeiT-T & DeiT-S & DeiT-B & Swin-T & Swin-S & Swin-B & QQP & SST-2 & MRPC & \begin{tabular}[c]{@{}c@{}}WikiText2\\ PPL\end{tabular} & \begin{tabular}[c]{@{}c@{}}WikiText2\\ PPL\end{tabular} & \begin{tabular}[c]{@{}c@{}}Lambada\\ Acc(\%)\end{tabular} \\ \midrule
Baseline & fp32 & 72.14 & 79.83 & 81.79 & 80.99 & 83.21 & 83.60 & 90.98 & 92.90 & 86.03 & 15.94 & 14.62 & 75.41 \\ \midrule
TATAA & \begin{tabular}[c]{@{}c@{}}int8+\\ bfloat16\end{tabular} & \begin{tabular}[c]{@{}c@{}}70.98\\ (-1.16)\end{tabular} & \begin{tabular}[c]{@{}c@{}}79.35\\ (-0.48)\end{tabular} & \begin{tabular}[c]{@{}c@{}}81.65\\ (-0.34)\end{tabular} & \begin{tabular}[c]{@{}c@{}}79.98\\ (-1.01)\end{tabular} & \begin{tabular}[c]{@{}c@{}}82.44\\ (-0.77)\end{tabular} & \begin{tabular}[c]{@{}c@{}}82.70\\ (-0.90)\end{tabular} & \begin{tabular}[c]{@{}c@{}}90.15\\ (-0.83)\end{tabular} & \begin{tabular}[c]{@{}c@{}}92.32\\ (-0.58)\end{tabular} & \begin{tabular}[c]{@{}c@{}}85.54\\ (-0.49)\end{tabular} & \begin{tabular}[c]{@{}c@{}}16.41\\ (+0.47)\end{tabular} & \begin{tabular}[c]{@{}c@{}}15.18\\ (+0.56)\end{tabular} & \begin{tabular}[c]{@{}c@{}}74.96\\ (-0.45)\end{tabular} \\ \bottomrule
\multicolumn{14}{l}{\footnotesize *Baseline models with pretrained \type{fp32} parameters are loaded from PyTorch or Hugging Face model hub.} \\
\multicolumn{14}{l}{\footnotesize $^{\dagger}$SmoothQuant \cite{xiao2023smoothquant} is applied for OPT-1.3B quantization.}
\end{tabular}
}
\label{tab:quant-eval}
\end{table}

Firstly, we evaluate the approximation techniques outlined in Section 5.2.2 to demonstrate that our \type{bfloat16} implementation of non-linear functions is precise and can therefore be used for complete model inference. \figref{fig:nonlinear_funtion_error} shows the errors for the inversed square root, pade $tanh$, as well as the funtion-level GELU approximation. We did not evaluate the power of 2 approximation as it has been widely utilized in SoftMax hardware and demonstrated to produce negligible errors \cite{DBLP:journals/corr/abs-2103-09301}. In the given input range, the overall RMSEs of approximations are $1.90 \times 10^{-3}$, $1.52 \times 10^{-2}$, and $1.97 \times 10^{-3}$ for the two methods and the GELU activation function, respectively, demonstrating our selected approximations for non-linear functions are sufficient.

Using the \type{int8} + \type{bfloat16} post-training quantization (PTQ) method, we evaluate several transformer models on a range of tasks, simulating model accuracy through PyTorch-based quantization codes. The calibration dataset is generated by randomly sampled a very small size of training set ($16\sim 128$ in our setups). \tabref{tab:quant-eval} presents the inference performance for ViT\footnote{Three scales of DeiT and Swin Transformer, -T, -S, -B refer to Tiny, Small and Base, respectively.}, BERT, and GPT-2 models, with classification and text generation tasks. The drop in accuracy among all evaluation tasks is negligible from $0.34\%$ to $1.16\%$, demonstrating that the \aname PTQ scheme is available for flexible transformer acceleration without the need for retraining overhead. As illustrated above, static PTQ is applied in \aname, and in the current work we can deploy other existing PTQ approaches like SmoothQuant \cite{xiao2023smoothquant} and FQ-ViT \cite{lin2022fqvit}, thanks to the general support in \aname framework. With the on-chip quantization \& layout conversion module, \aname can efficiently deploy quantized matrix multiplication and non-linear functions with appropriate data format and layout, as long as the framework gets static quantization scaling factors, according to \eqnref{eq:quant}.

\subsection{Hardware Utilization}

\begin{table}[tbp]
\centering
\caption{Hardware Utilization of the Proposed \aname Processing Core}
\label{tab:hw-utl}
\resizebox{0.7\linewidth}{!}{
\begin{tabular}{@{}ccccc@{}}
\toprule
\multirow{2}{*}{Components} & \multicolumn{4}{c}{FPGA Utilization} \\ \cmidrule(l){2-5} 
 & LUT & FF & BRAM & DSP \\ \midrule
DMPUs & 60117 ($86.8\%$) & 85035 ($85.4\%$) & 0 ($0.0\%$) & 512 ($94.1\%$) \\
Register Files & 2240 ($3.2\%$) & 4333 ($4.4\%$) & 78.5 ($54.0\%$) & 0 ($0.0\%$)\\
Dual-mode Buffers & 280 ($0.4\%$) & 224 ($0.2\%$) & 60 ($41.2\%$) & 0 ($0.0\%$)\\
Quantization Layout Convert & 6558 ($9.5\%$) & 9899 ($9.9\%$) & 7 ($4.8\%$) & 32 ($5.9\%$) \\
Controller Misc & 87 ($0.1\%$) & 80 ($0.1\%$) & 0 ($0.0\%$) & 0 ($0.0\%$) \\ \midrule
One TATAA Core Total & 69282 & 99571 & 145.5 & 544 \\ \bottomrule
\end{tabular}
}
\end{table}

\tabref{tab:hw-utl} presents the hardware utilization on FPGA based on the selected configuration. It can be concluded that the DMPUs dominate the resources cost, and other overhead units like quantization and transpose are relatively small. In detail, the proposed DMPUs cost $86.8\%$ LUTs, $85.4\%$ FFs, and $94.1\%$ DSPs for FPGA resources. Due to the SIMD approach in the \type{bfloat16} mode, the controller overhead is minimal because the dataflow is shared between all FPUs.

\begin{figure}[tbp]
    \centering
    \includegraphics[width=\linewidth]{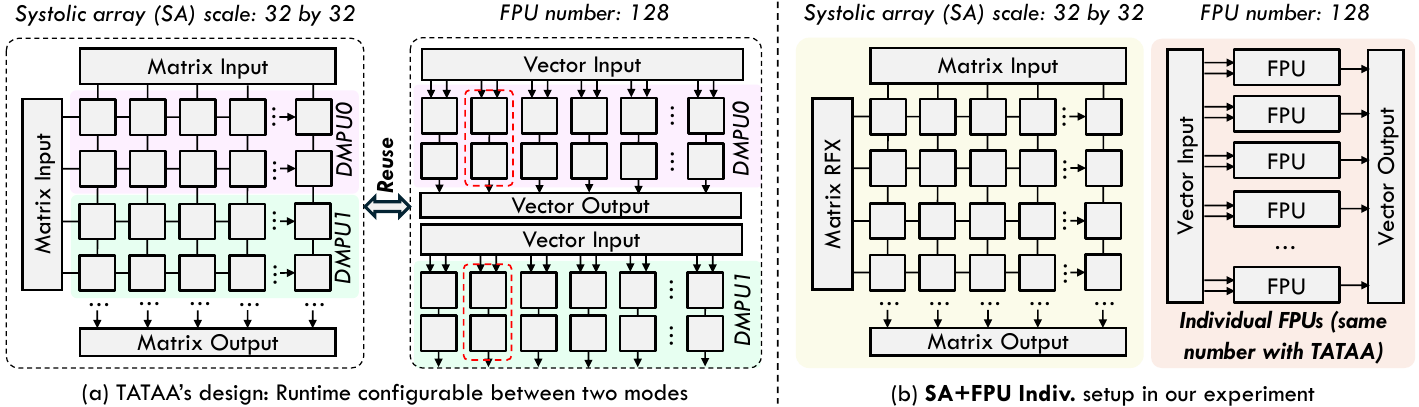}
    \caption{Experiment setup of comparing \aname (reusing same hardware for \textit{MatMul} and \type{bfloat16} operations) and traditional implementation (individual systolic array and FPU, \textbf{SA+FPU Indiv.} in abbreviation). The systolic array scale and the number of FPUs are the same for fair comparison.}
    \label{fig:indiv-setup}
\end{figure}

\begin{figure}
    \centering
    \includegraphics[width=0.69\linewidth]{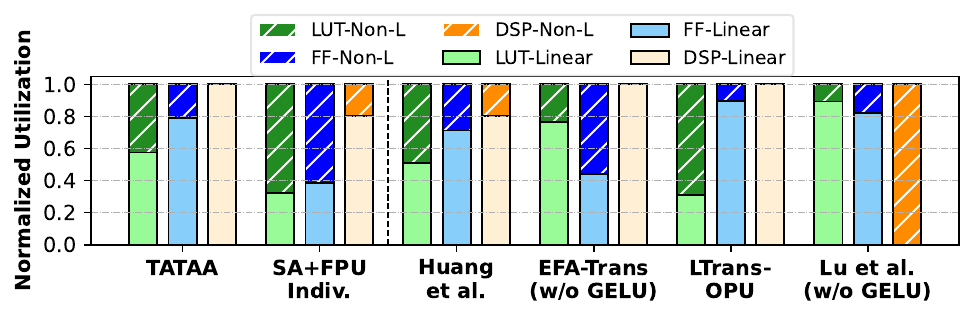}
    \caption{Normalized hardware utilization on FPGA for the proposed \aname and other related works, in terms of linear \textit{MatMul} units (-L) and non-linear functions overhead (-NL) based on three types of resources (LUT, FF, DSP).}
    \label{fig:hw-utl}
\end{figure}

We provide a detailed breakdown of utilization, separating integer linear units from the overhead needed to support non-linear operations, and the experiment setups are illustrated in \figref{fig:indiv-setup}, named as \textbf{SA+FPU Indiv.} design. For \aname and \textbf{SA+FPU Indiv.} setup, \figref{fig:hw-utl} presents the normalized utilization in terms of hardware units for both linear layers (\textit{MatMul}) and non-linear functions (indicated with shadows). We compare our \aname architecture with a design that utilizes individual integer systolic arrays and FPUs without reuse, on the same scales ($32$ by $32$ array and $128$ lane SIMD FPUs), as illustrated on the left side of \figref{fig:hw-utl} (\textbf{SA+FPU Indiv.}). Such a comparison indicates that the reuse scheme drastically reduces hardware costs for non-linear functions. Additionally, we present other related FPGA-based accelerators utilization of linear versus non-linear functions based on their reported results \cite{huang2023integer, yang2022efa, ltransopu, swat, lu2020hardware}. Our \aname architecture exhibits comparable overhead across three resource types, with only $10.5\%$ FFs, and no DSPs overhead especially. As exceptions, EFA-Trans \cite{yang2022efa} reports linear operation units including SoftMax, and in the work by Lu et al. \cite{lu2020hardware}, where LUTs are employed for linear computations and the DSP overhead for non-linear operations reaches $100\%$.

\subsection{\aname Runtime Analysis}

\begin{figure}
    \includegraphics[width=\linewidth]{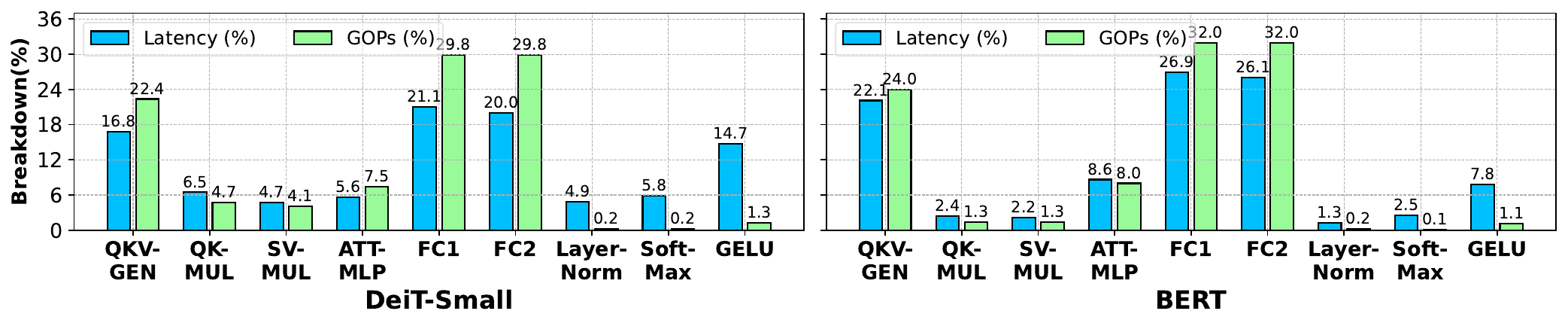}
    \caption{Layerwise latency \& computation workload size breakdown in DeiT-Small and BERT, during \aname runtime.}
    \label{fig:lat-gops}
\end{figure}

We choose DeiT-Small (DeiT-S) with batch size $16$ and BERT with sequence length $128$ \& batch size $32$ to measure layer latency with its workload size \figref{fig:lat-gops}. The linear \textit{MatMul} layers (such as QKV-GEN, QK-MUL, SV-MUL, etc.) are the primary contributors to the transformer workload when measured in terms of GOP (giga operations) or GFLOP (giga floating-point operations), and therefore heavily influence total latency. Among these linear layers, QK-MUL and SV-MUL are slightly less efficient, exhibiting a smaller workload-latency ratio compared to other MLP layers such as QKV-GEN. This inefficiency is attributed to the fact that the multi-head attention mechanism reduces the matrix sizes in each \textit{MatMul}, which in turn leads to less data reuse in the output-stationary dataflow. Such an analysis shows the optimization space in the future to dynamically optimize the different size of workloads. In addition, despite the non-linear functions having significantly smaller workloads compared to the linear \textit{MatMul} layers, they nevertheless contribute significantly to latency, accounting for approximately $25\%$ of the total end-to-end inference time. Hence, the implementation of non-linear functions is crucial for both model performance and hardware efficiency, as is also proved in some previous studies \cite{softermax}. Our flexible and adaptable framework for compiling these functions offers an optimization potential for new transformer models that utilize various non-linear functions.

\begin{figure}[tbp]
    \centering
    \includegraphics[width=0.56\linewidth]{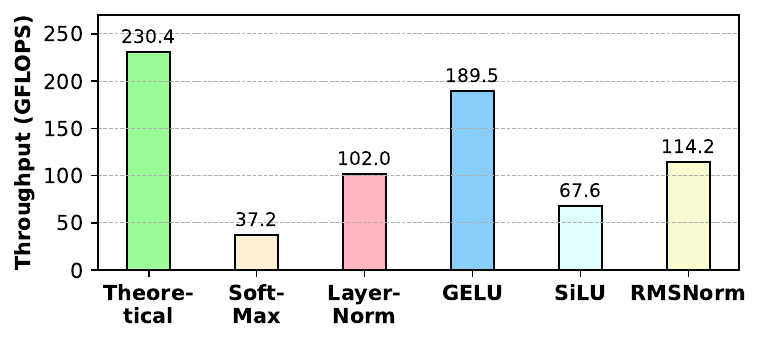}
    \caption{Evaluation of the selected non-linear functions in various transformer models. The throughput (GFLOPS) is measured on hardware runtime. The SoftMax, LayerNorm and GELU functions are based on BERT model, while the SiLU is used in ChatGLM and RMSNorm is based on Llama-7B.}
    \label{fig:eval-nonl-gflops}
\end{figure}

\figref{fig:eval-nonl-gflops} presents the non-linear functions throughput based on \type{bfloat16} basic operations (giga floating-point operations per second, GFLOPS), over several selected transformer models. Since the proposed \aname architecture supports full pipeline between basic operations (\textit{fpmul}, \textit{fpadd}, and \textit{fpapp}), the theoretical maximum throughput considering the computation resources (i.e., FPU number in \aname architecture), can be calculated by \eqnref{eq:eval-gflops}:

\begin{equation}
    GFLOPS_{theo} = K \cdot N \cdot W \cdot freq
    \label{eq:eval-gflops}
\end{equation}
where the $K \cdot N \cdot W$ refer to the number of FPUs (128 in our setup). In our evaluation setup, the throughput $GFLOPS_{theo} = 230.40$. Among all the test functions, our \aname can reach to maximum $189.45$ GFLOPS in GELU function of BERT model. This is because the memory-bound nature of these \type{bfloat16}-based non-linear functions. Still, our compilation framework can reach $82.2\%$ maximum throughput and leave an optimization space for compiler in the future. For instance, the SoftMax function requires accessing external memory several times due to the data dependency as illustrated in the compilation steps \figref{fig:compiler-nonl}, thus causing lower throughput as the compiled operations are highly memory-bound. The users can deploy more efficient SoftMax schemes to improve it, like Flash-Attention \cite{dao2022flashattentionfastmemoryefficientexact} which significantly reduces the memory I/O.

\begin{table}[tbp]
\centering
\caption{Normalized non-linear functions latency for \aname and related works for non-linear functions implementation.}
\resizebox{0.64\linewidth}{!}{
\begin{tabular}{@{}cccccc@{}}
\toprule
\multirow{3}{*}{\begin{tabular}[c]{@{}c@{}}Non-linear\\ Implementation\end{tabular}} & \multicolumn{4}{c}{Latency (cycles per element)} & \multirow{3}{*}{\begin{tabular}[c]{@{}c@{}}DSP\\ overhead\end{tabular}} \\ \cmidrule(lr){2-5}
 & \multirow{2}{*}{SoftMax} & \multirow{2}{*}{LayerNorm} & \multirow{2}{*}{GELU} & \multirow{2}{*}{Total} &  \\
 &  &  &  &  &  \\ \midrule
\multirow{2}{*}{\begin{tabular}[c]{@{}c@{}}NPE-1024\cite{khannpe}\\ (\type{fxp} Vector Unit)\end{tabular}} & \multirow{2}{*}{2.53} & \multirow{2}{*}{9.94} & \multirow{2}{*}{0.75} & \multirow{2}{*}{13.22} & \multirow{2}{*}{1.58\%} \\
 &  &  &  &  &  \\ \midrule
\multirow{2}{*}{\begin{tabular}[c]{@{}c@{}}Huang et al.\cite{huang2023integer}\\ (\type{fxp} Special Unit)\end{tabular}} & \multirow{2}{*}{1.59} & \multirow{2}{*}{2.40} & \multirow{2}{*}{-} & \multirow{2}{*}{-} & \multirow{2}{*}{18.5\%} \\
 &  &  &  &  &  \\ \midrule
\multirow{2}{*}{\begin{tabular}[c]{@{}c@{}}Chen el al.\cite{chenllm}\\ (\type{fp} Special Unit)\end{tabular}} & \multirow{2}{*}{5.14} & \multirow{2}{*}{0.66} & \multirow{2}{*}{0.13} & \multirow{2}{*}{5.92} & \multirow{2}{*}{21.0\%} \\
 &  &  &  &  &  \\ \midrule
\multirow{2}{*}{\begin{tabular}[c]{@{}c@{}}TATAA (\type{bfloat16}\\ Transformable Architecture)\end{tabular}} & \multirow{2}{*}{0.50} & \multirow{2}{*}{0.51} & \multirow{2}{*}{0.39} & \multirow{2}{*}{\begin{tabular}[c]{@{}c@{}}1.39\\ (4.25$\times$)\end{tabular}} & \multirow{2}{*}{0\%} \\
 &  &  &  &  &  \\ \bottomrule
\end{tabular}
}
\label{tab:non-linear-lat}
\end{table}

\begin{figure}[tbp]
    \centering
    \includegraphics[width=\linewidth]{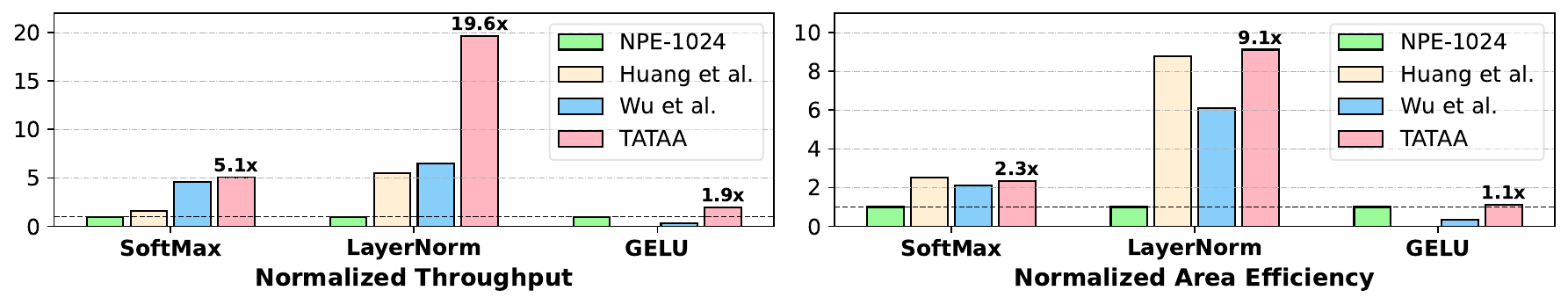}
    \caption{Comparison of normalized throughput and normalized area efficiency between \aname and related works. The area efficiency is measured by DSP utilization.}
    \label{fig:eval-nonl-comp}
\end{figure}

We further compare the latency of non-linear functions in \aname with various related studies that have documented their evaluations of nonlinear functions, as shown in \tabref{tab:non-linear-lat}. Since the token lengths and model scales in these acceleration works are different, we normalize the latency to cycles per element, as the same setup in \cite{khannpe}. The proposed \aname achieves significantly lower latency in terms of SoftMax and LayerNorm function, because the transformable architecture is able to utilize all the processing units for non-linear functions, boosting the theoretical floating-point operations throughput. In terms of GELU, since \aname only applies naive approximation to demonstrate our flexibility, the final latency is not good as Chen et al. \cite{chenllm}. But in general, our total latency for non-linear functions still outperforms Chen et al. by $4.25\times$, without any computational resources overhead. In addition, we compare the proposed implementation with previous works in terms of throughput and area efficiency (throughput/DSP blocks), as shown in \figref{fig:eval-nonl-comp}. The results show that \aname also achieves higher throughput and comparable area efficiency in DSPs (reach $19.6\times$ and $9.1\times$ higher than the baseline NPE-1024 design \cite{khannpe}), while other works may still cost more LUTs or FFs for non-linear functions. Compared to our previous study \cite{wu2024case} which fuses \type{fp32} and 8-bit block floating point (\type{bfp8}) with the similar idea, \aname achieves higher throughput improvement since we are targeting a cheaper data format (\type{int8} and \type{bfloat16}) and a more efficient hardware reuse scheme. All in all, the key benefit of \aname is the potential for further optimization of efficiency through compilation of emerging non-linear functions, a feature that is absent in those accelerators with fixed and specific units.

\subsection{Resource Efficiency}

\begin{figure}[tbp]
    \centering
    \includegraphics[width=0.6\linewidth]{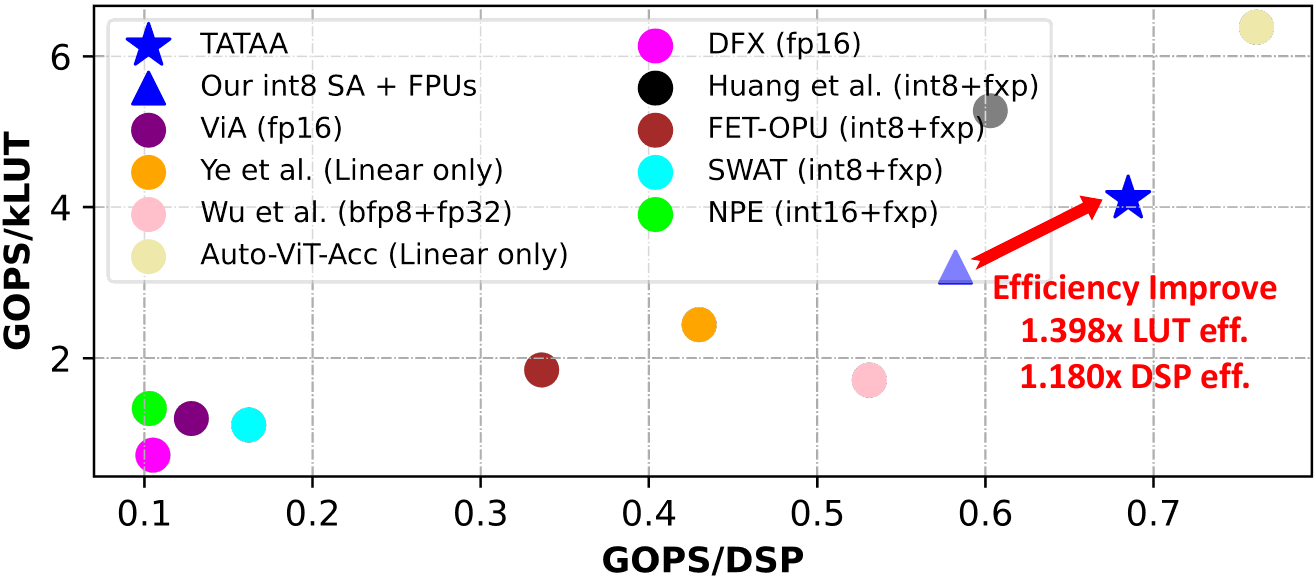}
    \caption{Resource efficiency in terms of GOPS/DSP and GOPS/kLUT, comparing \aname with several FPGA-based acceleration frameworks}
    \label{fig:area-eff-linear}
\end{figure}

As one of the key contributions in \aname, we reuse the same integer hardware units for non-linear functions deployment, saving the hardware overhead and thus improving the resource/area efficiency. Firstly, we evaluate end-to-end throughput (giga operations per second, GOPS) of \aname and normalize the throughput into resource to obtain resource efficiency, in terms of both LUT and DSP. For comparison, we selected related FPGA-based accelerators for transformers, calculating their efficiency based on the utilization results. The selected models have various data format setups, e.g., DFX \cite{hong2022dfx} implements \type{fp16} for all operations while Auto-ViT-Acc \cite{lit2022auto} only targets linear \textit{MatMul} in integer. To benchmark our transformable architecture, we also implement the \textbf{SA+FPU Indiv.} design, the opposite setup compared to \aname, as shown in \figref{fig:indiv-setup}.

\figref{fig:area-eff-linear} presents the resource efficiency results based on the selected implementations, in terms of end-to-end GOPS per DSP (GOPS/DSP) and GOPS per kilo-LUT (GOPS/kLUT). Compared with our own baseline (\textbf{SA+FPU Indiv.}), \aname achieves $1.28\times$ LUT efficiency and $1.18\times$ DSP efficiency, due to the hardware reusing scheme in the proposed transformable architecture. Compared to other related works, our \type{int8} + \type{bfloat16} approach is still compatible with smaller workload in linear layers. Although some previous works demonstrate superior resource efficiency in terms of LUT or DSP, it is important to mention that they either do not support full inference on hardware (Auto-ViT-Acc \cite{lit2022auto}) or employ more aggressive approximations of non-linear functions using fixed-point format (Huang et al. \cite{huang2023integer}), lacking the PTQ retrain-free benefit that is implemented in our \aname. Besides, Wu et. al \cite{wu2024case} proposed to fusing block floating-point \type{bfp8} and \type{fp32} formats in the same architecture which is similar to \aname. However, due to the higher hardware cost for block-wise operation and \type{fp32}, our \aname achieves around $2.25\times$ and $1.31\times$ higher efficiency for GOPS/kLUT and GOPS/DSP.

\subsection{Systematic Comparison with Related Studies}

\begin{table}[tbp]
\caption{Hardware Performance Comparison with Relative FPGA-based Accelerators for Transformer Models}
\resizebox{\linewidth}{!}{
\begin{threeparttable}
\begin{tabular}{@{}cccccccccccccc@{}}
\toprule
\multirow{2}{*}{Work} & \multirow{2}{*}{\begin{tabular}[c]{@{}c@{}}Data\\ Formats\tnote{$\ddagger$}\end{tabular}} & \multirow{2}{*}{\begin{tabular}[c]{@{}c@{}}End2end\\ Support\end{tabular}} & \multirow{2}{*}{\begin{tabular}[c]{@{}c@{}}FPGA\\ Platform\end{tabular}} & \multicolumn{4}{c}{FPGA Utilization} & \multirow{2}{*}{\begin{tabular}[c]{@{}c@{}}Freq.\\ (MHz)\end{tabular}} & \multirow{2}{*}{\begin{tabular}[c]{@{}c@{}}Power\\ (W)\end{tabular}} & \multirow{2}{*}{\begin{tabular}[c]{@{}c@{}}Eval.\\ Models\end{tabular}} & \multirow{2}{*}{\begin{tabular}[c]{@{}c@{}}Throughput\\ Inf./sec\tnote{$\mathsection$}\end{tabular}} & \multirow{2}{*}{\begin{tabular}[c]{@{}c@{}}Throughput\\ (GOPS)\end{tabular}} & \multirow{2}{*}{\begin{tabular}[c]{@{}c@{}}DSP\\ Efficiency\end{tabular}} \\ \cmidrule(lr){5-8}
 &  &  &  & LUT(k) & FF(k) & BRAM & DSP &  &  &  &  &  &  \\ \midrule
\multirow{2}{*}{\begin{tabular}[c]{@{}c@{}}Auto-ViT-\\ Acc \cite{lit2022auto}\end{tabular}} & \multirow{2}{*}{fxp, fp32} & \multirow{2}{*}{No} & \multirow{2}{*}{ZCU102} & \multirow{2}{*}{185.0} & \multirow{2}{*}{-} & \multirow{2}{*}{-} & \multirow{2}{*}{1552} & \multirow{2}{*}{150} & \multirow{2}{*}{9.6} & DeiT-S & 99.7 & 907.8 & 0.585 \\
 &  &  &  &  &  &  &  &  &  & DeiT-B & 34.0 & 1181.5 & 0.761 \\ \midrule
\multirow{2}{*}{\begin{tabular}[c]{@{}c@{}}Huang\\ et al. \cite{huang2023integer}\end{tabular}} & \multirow{2}{*}{int8, int8} & \multirow{2}{*}{Yes} & \multirow{2}{*}{ZCU102} & \multirow{2}{*}{144.5} & \multirow{2}{*}{168.0} & \multirow{2}{*}{648} & \multirow{2}{*}{1268} & \multirow{2}{*}{300} & \multirow{2}{*}{29.6} & ViT-S & 89.7 & 762.7 & 0.601 \\
 &  &  &  &  &  &  &  &  &  & ViT-T & 245.3 & 616.1 & 0.486 \\ \midrule
\multirow{2}{*}{HPTA\cite{hpta}} & \multirow{2}{*}{int8, int8} & \multirow{2}{*}{Yes} & \multirow{2}{*}{ZCU102} & \multirow{2}{*}{209.9} & \multirow{2}{*}{368.4} & \multirow{2}{*}{345} & \multirow{2}{*}{2307} & \multirow{2}{*}{200} & \multirow{2}{*}{20.0} & BERT & 81.9 & - & 0.035\tnote{$\dagger$} \\
 &  &  &  &  &  &  &  &  &  & Swin-T & 148.8 & - & 0.065\tnote{$\dagger$} \\ \midrule
NPE \cite{khannpe} & int16, fxp & Yes & VCU118 & 192.4 & 351.1 & 369 & 2020 & 200 & 20.0 & BERT & 36.8 & - & 0.018\tnote{$\dagger$} \\ \midrule
FTRANS \cite{ftrans} & fp16, fp32 & Yes & VCU118 & 451.1 & 506.6 & - & 6531 & - & - & RoBERTa & 94.25 & - & 0.014 \\ \midrule
ViA \cite{wang2022via} & fp16, fp16 & Yes & Alveo U50 & 258.0 & 257.0 & 1022 & 2420 & 300 & 39.0 & Swin-T & - & 309.6 & 0.128 \\ \midrule
SWAT \cite{swat} & int8, fxp & Yes & Alveo U50 & 271.0 & - & 609.5 & 1863 & 200 & 14.4 & Swin-T & - & 301.9 & 0.162 \\ \midrule
\multirow{2}{*}{ME-ViT\cite{marino2023me}} & \multirow{2}{*}{int8, fxp} & \multirow{2}{*}{Yes} & \multirow{2}{*}{Alveo U200} & \multirow{2}{*}{192.0} & \multirow{2}{*}{132.0} & \multirow{2}{*}{288} & \multirow{2}{*}{1024} & \multirow{2}{*}{300} & \multirow{2}{*}{9.3} & DeiT-B & 23.9 & - & 0.0233\tnote{$\dagger$} \\
 &  &  &  &  &  &  &  &  &  & DeiT-S & 41.7 & - & 0.0407\tnote{$\dagger$} \\ \midrule
DFX \cite{hong2022dfx} & fp16, fp16 & Yes & Alveo U280 & 520.0 & 1107.0 & 1192 & 3533 & 200 & - & GPT-2 & 0.361 & 185.6 & 0.0001\tnote{$\dagger$} \\ \midrule
Ye et al.\cite{yeetal} & int8, fxp & No & Alveo U250 & 736.0 & - & 1781 & 4189 & 300 & - & - & - & 1800.0 & 0.430 \\ \midrule
\multirow{3}{*}{FET-OPU\cite{bai2023fet}} & \multirow{3}{*}{int8, fxp} & \multirow{3}{*}{Yes} & \multirow{3}{*}{Alveo U280} & \multirow{3}{*}{886.8} & \multirow{3}{*}{716.6} & \multirow{3}{*}{1357} & \multirow{3}{*}{4864} & \multirow{3}{*}{200} & \multirow{3}{*}{7.4} & DeiT-B & 71.8 & 1264.6 & 0.0148\tnote{$\dagger$} \\
 &  &  &  &  &  &  &  &  &  & BERT & 146.6 & 1635.8 & 0.0301\tnote{$\dagger$} \\
 &  &  &  &  &  &  &  &  &  & Swin-T & 124.1 & 1070.1 & 0.0256\tnote{$\dagger$} \\ \midrule
\multirow{10}{*}{\aname} & \multirow{10}{*}{\begin{tabular}[c]{@{}c@{}}int8,\\ bfloat16\end{tabular}} & \multirow{10}{*}{Yes} & \multirow{10}{*}{Alveo U280} & \multirow{10}{*}{724.9} & \multirow{10}{*}{1154.9} & \multirow{10}{*}{1472} & \multirow{10}{*}{4352} & \multirow{10}{*}{225} & \multirow{10}{*}{10.8} & \multirow{2}{*}{DeiT-S} & \multirow{2}{*}{218.6} & \multirow{2}{*}{2836.2\tnote{*}} & 0.626 \\
 &  &  &  &  &  &  &  &  &  &  &  &  & 0.0502\tnote{$\dagger$} \\ \cmidrule(l){11-14} 
 &  &  &  &  &  &  &  &  &  & \multirow{2}{*}{DeiT-B} & \multirow{2}{*}{67.6} & \multirow{2}{*}{2796.5} & 0.643 \\
 &  &  &  &  &  &  &  &  &  &  &  &  & 0.0156\tnote{$\dagger$} \\ \cmidrule(l){11-14} 
 &  &  &  &  &  &  &  &  &  & \multirow{2}{*}{BERT} & \multirow{2}{*}{116.8} & \multirow{2}{*}{2935.2} & 0.674 \\
 &  &  &  &  &  &  &  &  &  &  &  &  & 0.0269\tnote{$\dagger$} \\ \cmidrule(l){11-14} 
 &  &  &  &  &  &  &  &  &  & \multirow{2}{*}{Swin-T} & \multirow{2}{*}{179.7} & \multirow{2}{*}{2512.3} & 0.685 \\
 &  &  &  &  &  &  &  &  &  &  &  &  & 0.0587\tnote{$\dagger$} \\ \cmidrule(l){11-14} 
 &  &  &  &  &  &  &  &  &  & \multirow{2}{*}{GPT-2} & \multirow{2}{*}{7.9} & \multirow{2}{*}{2579.4} & 0.593 \\
 &  &  &  &  &  &  &  &  &  &  &  &  &  0.0018\tnote{$\dagger$} \\ \bottomrule
\end{tabular}
\begin{tablenotes}[]
\footnotesize
\item[$\ddagger$] Data formats for linear \textit{MatMul} (the former one) and non-linear functions (the latter one). \type{fxp} refers to fixed-point numbers.
\item[$\mathsection$] Inference per second (Inf./sec) measures how many end-to-end images or sequences can be processed through hardware in one second.
\item[$*$] We clarify that in our work, the total operation is obtained by doubling the MAC operation. Some previous work may directly report MACs as throughput.
\item[$\dagger$] Results with inference/sec/DSP are marked with symbols $\dagger$, while those based on GOPS/DSP are indicated separately. We provide both results for \aname.
\end{tablenotes}
\end{threeparttable}
 }
\label{tab:comp-related}
\end{table}

We summarize and compare other related FPGA-based transformer accelerators with our \aname FPGA prototype with respect to throughput and resource efficiency, as illustrated in \tabref{tab:comp-related}. In our setup, the ViT models (DeiT, Swin) are evaluated on ImageNet with a batch size of $16$, the BERT sequence length is fixed at $128$, and the GPT-2 with 345M parameters is under $512$ sequence length. We only measure the pre-fill stage for the GPT-2 inference. For other related works, we scaled their results to match ours if they set different sequence lengths or batch sizes, for a fair comparison. Our \aname achieves up to $2836.2$ GOPS with an end-to-end acceleration rate of maximum $218.6$ inference per second (Inf./sec) for vision models, while reaching $2579.4\sim 2935.2$ GOPS in language models. We also report throughput efficiency by normalizing the total throughput with computational resources (DSPs in FPGA), offering results in terms of GOPS/DSP or Inf./sec/DSP to facilitate comparisons across all related works. Compared to other accelerators, \aname obtains higher resource efficiency by factors ranging from $1.13\times$ to $2.29\times$ in terms of Inf./sec/DSP among all small-scale models, except in certain cases where end-to-end support is lacking (e.g., Auto-ViT-Acc \cite{lit2022auto}), or in some dedicated optimization work (e.g., FET-OPU \cite{bai2023fet} achieves $\sim 0.003$ higher efficiency than \aname). Nevertheless, the most important benefit of \aname is to support general and flexible non-linear functions, unlike other works deploying specific support for these functions.

We acknowledge that there are other leading acceleration frameworks that surpass \aname in terms of throughput and area efficiency. For instance, the work by Chen et al. \cite{chenllm} introduced a complete pipeline model using a spatial accelerator on FPGAs, with each block layer allocated to a separate hardware module, maintaining adequate buffer capacities for the pipeline. This configuration significantly enhances throughput by minimizing memory I/O operations, but as transformer models expand, mapping and compiling such a comprehensive pipeline approach becomes increasingly complex. Our strategy, in contrast, focuses on developing a general-purpose and broadly applicable accelerator. SSR's proposal by Zhuang et al. \cite{zhuangssr} involves implementing a spatial and temporal hybrid design on Versal ACAP devices, where \type{MatMul} operations utilize the AI Engine of AMD FPGAs. Directly comparing their framework's efficiency with ours wouldn't be equitable. Nonetheless, the primary innovation in \aname is presenting a novel methodology for reusing the same hardware across varied operations within transformer models. Thus, \aname can operate independently of spatial or temporal architectures in transformer accelerators. There is significant potential to further refine pipeline strategies among different \aname cores to facilitate the \type{MatMul} and \type{bfloat16} pipeline, and we intend to explore this potential in the future, inspired by these related works.

\begin{figure}[tbp]
    \centering
    \includegraphics[width=0.64\linewidth]{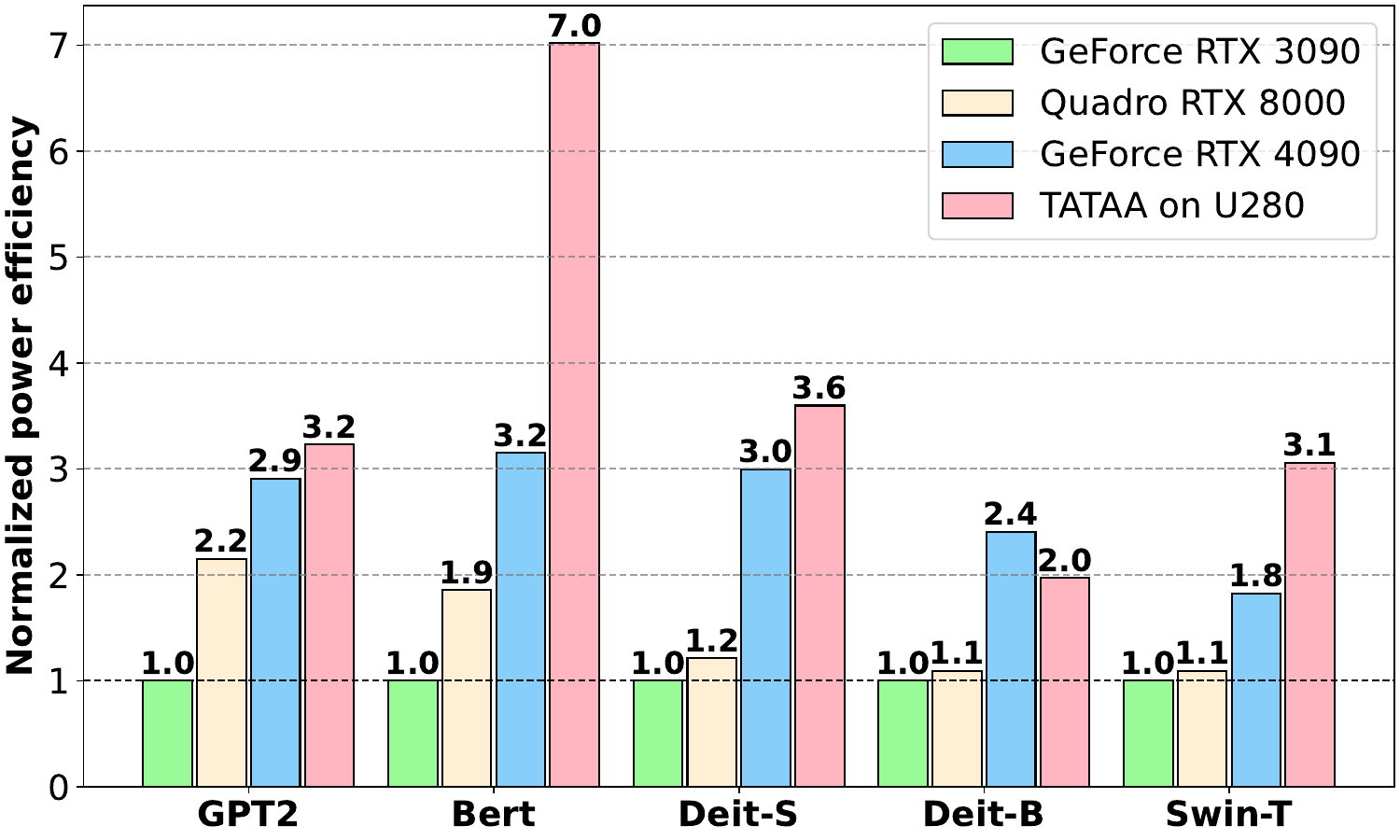}
    \caption{Nomalized power efficiency (Inf./sec/W) comparison between TATAA and GPU implementations.}
    \label{fig:pwr-eff}
\end{figure}

\subsection{Power Efficiency versus GPUs}
    Power efficiency, as an important highlight for the TATAA to be deployed on U280, shows prominent features of FPGA over modern GPUs. We used Xilinx RunTime(XRT) for hardware execution, and Vitis embeded power profile with Xilinx Board Utility (xbutil) for computing power measurements. NVIDIA system management interface (nvidia-smi) is used for measuring GPU power on Quadro RTX8000, GeForce RTX3090, and GeForce RTX4090.

We evaluate the internal power consumption of the TATAA framework, assessing the power efficiency (inferences per watt) for each model. Subsequently, we measured the model execution latency on various GPUs using consistent batch sizes and sequence lengths as in the TATAA framework, thereby computing the power efficiency (Inf./sec/W). As depicted in Figure \ref{fig:pwr-eff}, our TATAA FPGA surpasses the performance of the RTX3090 and RTX8000 across all models, demonstrating a $1.10\times$ improvement over the RTX4090 in normalized power efficiency for GPT2. For smaller models, such as the small DeiT transformer, TATAA remains competitive, achieving comparable power efficiency in end-to-end throughput. While TATAA shows a slight degradation in efficiency on Deit-B compared to the RTX4090, it significantly outperforms other devices on the BERT model ($2.19\times$ more than 4090) when the sequence length of inputs is properly adapted to the dataflow requirements of the TATAA hardware. This emphasizes one of the critical advantages of deploying TATAA on FPGAs compared to GPUs, highlighting its potential for processing large models with superior power efficiency.

\section{Conclusions}
In this work, we have presented \aname, a programmable accelerators on FPGA for transformer models by using a novel transformable arithmetic architecture.  Using \aname, we demonstrate that both low-bitwidth integer (\type{int8}) and floating-point (\type{bfloat16}) operations can be implemented efficiently using the same underlying processing array hardware.  By transforming the array from systolic mode for \type{int8} matrix multiplication to SIMD-mode for vectorized \type{bfloat16} operations, we show that end-to-end acceleration of modern transformer models including both linear and non-linear functions can be achieved with state-of-the-art performance and efficiency.  In the future, we plan to explore more general FPGA implementations of \aname with more devices support (i.e., with or without HBM) and to enhance the flexibility of our compilation framework to accelerate future transformer models as they are being rapidly developed.
\section{Acknowledgements}

This work was supported in part by the Research Grants Council (RGC) of Hong Kong under the Research Impact Fund project R7003-21 and the Theme-based Research Scheme (TRS) Project T45-701-22-R. This work was supported by AI Chip Center for Emerging Smart Systems (ACCESS), sponsored by InnoHK funding, Hong Kong SAR.

\bibliographystyle{unsrt}  
\bibliography{refs}

\end{document}